\documentclass[12pt,preprint]{aastex}
\usepackage{epsfig}
\usepackage{lscape}
\usepackage{rotating}
\usepackage{natbib}
\newcommand{\ch}{\textit {Chandra}}
\newcommand{\cxo}{\textit {Chandra X-ray Observatory}}
\newcommand{\xmm}{\textit {XMM}}
\newcommand{\xmmn}{\textit {XMM-Newton}}
\newcommand{\suzaku}{\textit {Suzaku}}

\shorttitle{\ch\ observation of IRAS 19254--7245}
\shortauthors{Jia et al.}

\begin{document}
\title{A \ch\ observation of the ultraluminous infrared galaxy IRAS 19254--7245 (the Superantennae):\\
        X-ray emission from the Compton-thick AGN and the diffuse starburst}
\author{Jianjun Jia\altaffilmark{1}, Andrew Ptak\altaffilmark{2}, Timothy M. Heckman\altaffilmark{1}, Valentina Braito\altaffilmark{3}, \& James Reeves\altaffilmark{4}}
\affil{\altaffilmark{1}Department of Physics and Astronomy, Johns Hopkins University, Baltimore, MD 21218, USA\\
\altaffilmark{2}Goddard Space Flight Center, Greenbelt, MD 20771, USA\\
\altaffilmark{3}Leicester University, UK\\
\altaffilmark{4}Keele University, UK}

\begin{abstract}

We present a \ch\ observation of IRAS 19254--7245, a nearby ULIRG also known as {\it the Superantennae}. The high spatial resolution of \ch\ allows us to disentangle for the first time the diffuse starburst emission from the embedded Compton-thick AGN. The 2-10 keV spectrum of the AGN emission is fitted by a flat power-law ($\Gamma=1.3$) and a He-like Fe K$\alpha$ line with EW$\sim$1.5 keV, consistent with previous observations. The Fe K$\alpha$ line profile could be resolved as a blend of a neutral 6.4 keV line and an ionized 6.7 keV (He-like) or 6.9 keV (H-like) line. Variability is detected compared with the previous \xmmn\ and \suzaku\ observations, demonstrating the compact size of the iron line emission. We fit the spectrum of the galaxy-scale extended emission excluding the AGN and other bright point sources with a soft thermal component with $kT\sim 0.8$ keV. The luminosity of the extended emission is about one order of magnitude lower than that of the AGN. The basic physical and structural properties of the extended emission are fully consistent with a galactic wind being driven by the starburst (no contribution to the feedback by the AGN is required).  A candidate ultra-luminous X-ray source is detected 8\arcsec\ south of the southern nucleus. The 0.3-10 keV luminosity of this off-nuclear point source is $\sim 6\times 10^{40}$ erg s$^{-1}$ if the emission is isotropic and the source is associated with the Superantennae. 

\end{abstract}


\section{Introduction}

Ultra-Luminous Infrared Galaxies (ULIRGs) emit most of their total energy output in the far-infrared (FIR) domain, with a FIR luminosity greater than $10^{12} L_{\odot}$. Many ULIRGs are a composite of an active galactic nucleus (AGN) and a starburst (SB), with the violent star formation activity triggered by a galaxy collision or merger. The emission from the central AGN and the dense star-forming regions is obscured by dust and gas, and the fractional contributions to the luminosity by the AGN and SB are not well constrained. Obtaining better insights into the energetics of these systems is crucial to understand the interplay between supermassive black holes (SMBHs) and the formation and evolution of galaxies, which is one of the major unsolved problems in astrophysics.

X-ray observations of ULIRGs above 3 keV, performed with \xmmn\ \citep{2003MNRAS.343.1181F,2003A&A...398..107B} and \ch\ \citep{2003ApJ...592..782P,2005ApJ...633..664T} have proved to be a fundamental tool to disentangle the contribution of SB and AGN activity and to investigate the presence of hidden AGNs in these sources. Indeed some of the ULIRGs, classified as pure SBs based on optical and IR spectroscopy, show spectral properties typical of obscured AGNs only when observed in X-rays (i.e., NGC 6240: \citealt{1998MNRAS.297.1219I}; Arp 299: \citealt{2002ApJ...581L...9D}). Previous X-ray surveys of ULIRGs have also shown that thermal emission from a hot plasma with a rather constant temperature $kT\sim 0.7$ keV is always present \citep{2005ApJ...628..187G}. This emission dominates the X-ray spectra below 1 keV and it is likely associated with a nuclear or circumnuclear starburst. In particular, \ch\ observations of luminous and ultraluminous IR galaxies have proved to be essential to investigate the spatial structure and physical properties of this soft diffuse X-ray emission. Thanks to its high spatial resolution, \ch\ observations have shown that in ULIRGs, like in local starbursts, this soft diffuse emission extends far beyond the starburst region itself. This emission is primary evidence of the presence of galactic superwinds \citep{1985Natur.317...44C,1993ASSL..188..455H,2000MNRAS.314..511S}, which are believed to play a major role in the evolution of galaxies and metal enrichment of the intergalactic medium. However, these studies have also shown that ULIRGs are intrinsically faint X-ray sources, thus an X-ray follow up is feasible only for the very brightest ones.

Understanding the physical processes at work in ULIRGs is important since observational evidence shows that ULIRGs are generally advanced mergers of gas-rich galaxies, which are considered to be the origin of many massive elliptical and S0 galaxies. If indeed ULIRGs are forming spheroids, an AGN/SB association is naturally supported by the evidence that all spheroidal galaxies host quasar relics in the form of supermassive black holes (SMBHs), and the quasar stage could be a phase during the evolution of these systems. Support for this view has come from high-resolution spectroscopy of ULIRG mergers \citep{2001ApJ...563..527G}, whose dynamical properties have proved that they indeed are ``ellipticals in formation". In this framework high resolution simulations of galaxies mergers \citep{2006ApJ...652..864H,2005MNRAS.361..776S} have shown that galaxy mergers may play a fundamental role in supermassive black hole growth and galaxy evolution. Indeed the merger provides the inflow of gas which is likely to fuel both the starburst and, at some stages, an AGN. These simulations predict that for most of this phase the AGN is obscured, with the optically-visible AGN phase occurring in the late stages when the starburst and/or AGN essentially explosively drives out all the remaining interstellar material in the system. To test the predictions of the evolutionary sequence more robust SFR and AGN activity estimators are needed. This reinforces the importance of understanding the contributions of SB and AGN activity in known examples in which they coexist.

Like most ULIRGs, IRAS 19254--7245, at $z=0.062$, is a merger system of two gas-rich spiral galaxies, with the nuclei $\sim 9\arcsec$ (which corresponds to $\sim$11 kpc) apart from each other. Among these interacting ULIRGs, IRAS 19254--7245, with $L_{\rm FIR}=2\times 10^{12} L_{\odot}$, is a remarkable one. It is known to host a powerful SB with star formation rate (SFR) $\sim 80 {\rm M_{\odot}/yr}$ \citep{2003A&A...403..119B} and a heavily obscured AGN, which resembles in many aspects the prototypical type-II (obscured) quasar NGC 6240 \citep{1998MNRAS.297.1219I}. This source exhibits giant tidal tails extending to a distance of about 350 kpc triggered by a merger event reminiscent of the classical ``{\it Antennae}" seen in NGC 4038/4039. The southern galaxy is classified as a Seyfert 2 \citep{1991A&A...243..367M}, and the northern galaxy shows an optical spectrum typical of SB \citep{2003A&A...403..119B}.

Previous X-ray observations of IRAS 19254--7245 are strongly indicative that this ULIRG harbors both a powerful SB and a deeply buried high-luminosity AGN \citep{2000MNRAS.314..589P,2003A&A...398..107B}. A recent \suzaku\ observation has detected the intrinsic AGN emission in the 15-30 keV band and has confirmed the \xmmn\ detection of Fe-K$\alpha$ emission \citep{2009A&A...504...53B}. These data show that both neutral and ionized Fe-K$\alpha$ lines are present. In this scenario, the intrinsic 2-10 keV luminosity of IRAS 19254--7245 could be greater than $10^{44}~{\rm ergs~s^{-1}}$, dominated by the buried quasar. This picture has been confirmed by high signal to noise L band ($3-4\mu m$) \citep{2003ApJ...595L..17R} and Spitzer observations \citep{2008MNRAS.385L.130N} which also suggested that the main energy source in the mid-IR is an obscured AGN. The \xmmn\ and \suzaku\ observations have also shown that the soft emission is dominated by the starburst. The temperature $kT\sim 0.85~{\rm keV}$ and the luminosity of this soft component are consistent with the hot gas found in starburst galaxies and in the other ULIRGs observed so far. The \xmmn\ data show that, although the X-ray emission comes mostly from the southern nucleus containing the buried AGN, the source slightly extended. However, the \xmm\ PSF was too broad to allow the structure of the soft X-ray emission to be investigated in more detail. Furthermore, since the smallest possible extraction region with \xmmn\ includes the whole system, it was not possible to spatially disentangle the contribution of diffuse SB and AGN emission. Again despite the intrinsic high luminosity of the AGN in IRAS 19254--7245, it is heavily obscured and thus it may not outshine the starburst emission below 2 keV.

Here we present the analysis of three exposures of IRAS 19254--7245 by \ch\ with total exposure time of 75 ks. These are the first X-ray observations that allow us to directly disentangle the AGN and starburst. We present the \ch\ data reduction in Section \ref{data}, and we analyze and fit the spectra of the different parts of IRAS 19254--7245 in Section \ref{results}. We then discuss our results and draw conclusions in Section \ref{discussion}.

\section{Observations and data reduction \label{data}}

The \cxo\ observed IRAS 19254--7245 for three separated exposures with the back-illuminated chip S3 of the Advanced CCD Imaging Spectrometer (ACIS) instrument in a \texttt{FAINT} mode in December 2009 (see Table \ref{t:summary}). The data were reprocessed using the contributed script \texttt{chandra\_repro} of the CIAO software (ver. 4.3) and calibration database (CALDB 4.4.2) to create the new Level 2 event files. Sources were detected using the \texttt{wavdetect} tool.

\begin{figure}
\centering
\begin{tabular}{cc}
\epsfig{file=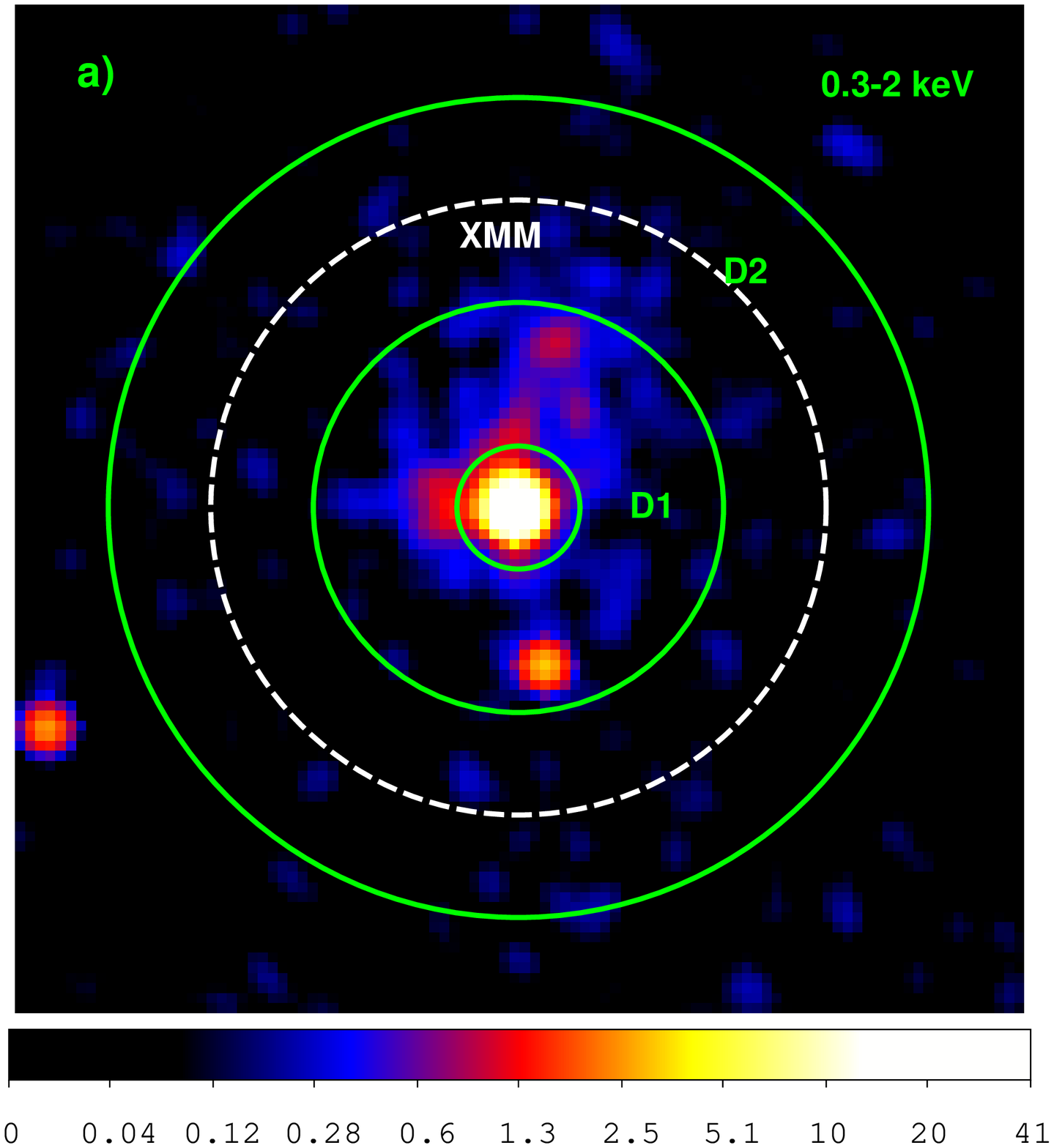,width=0.4\linewidth,angle=0,clip=} &
\epsfig{file=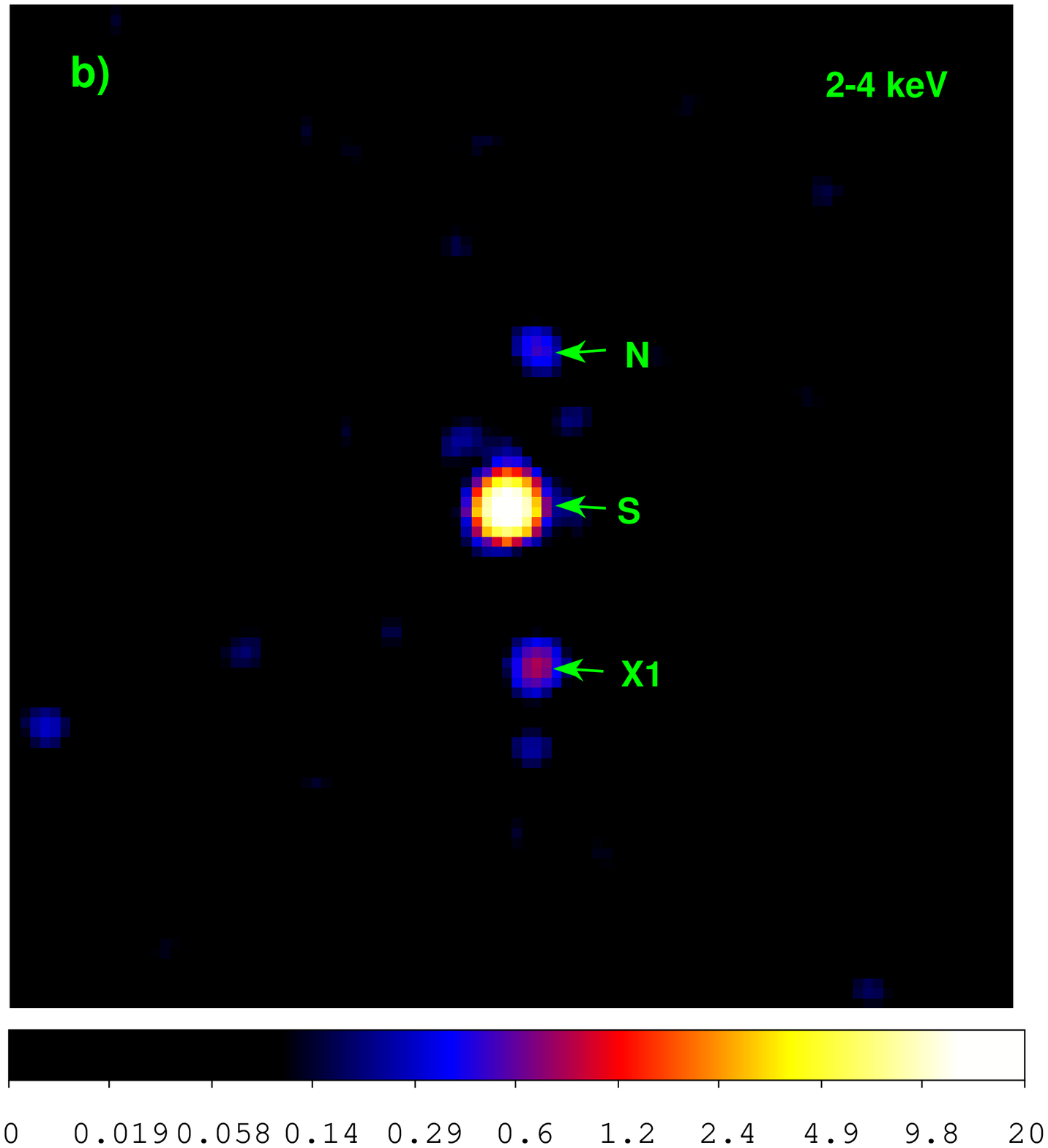,width=0.4\linewidth,angle=0,clip=} \\
\epsfig{file=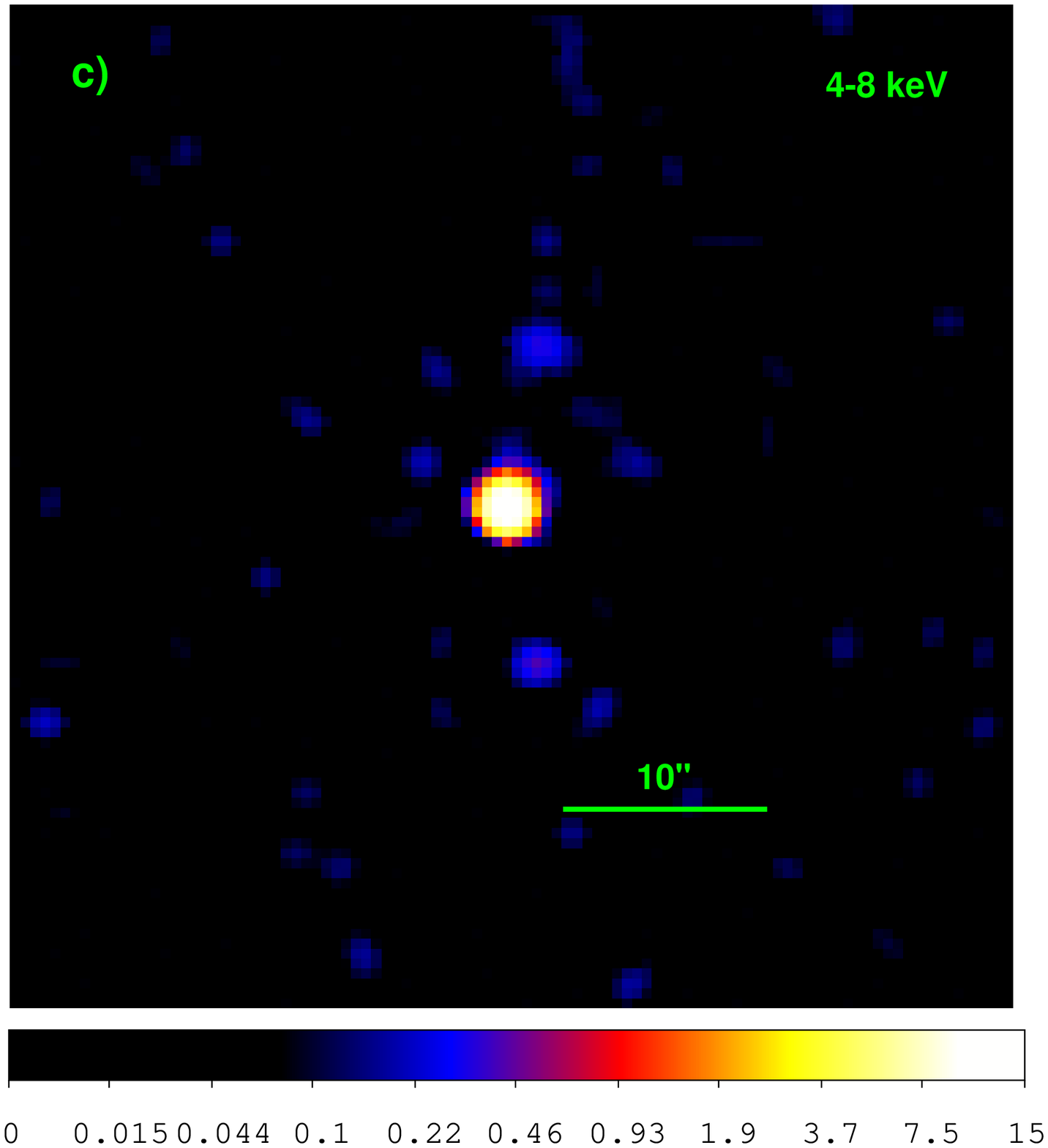,width=0.4\linewidth,angle=0,clip=} &
\epsfig{file=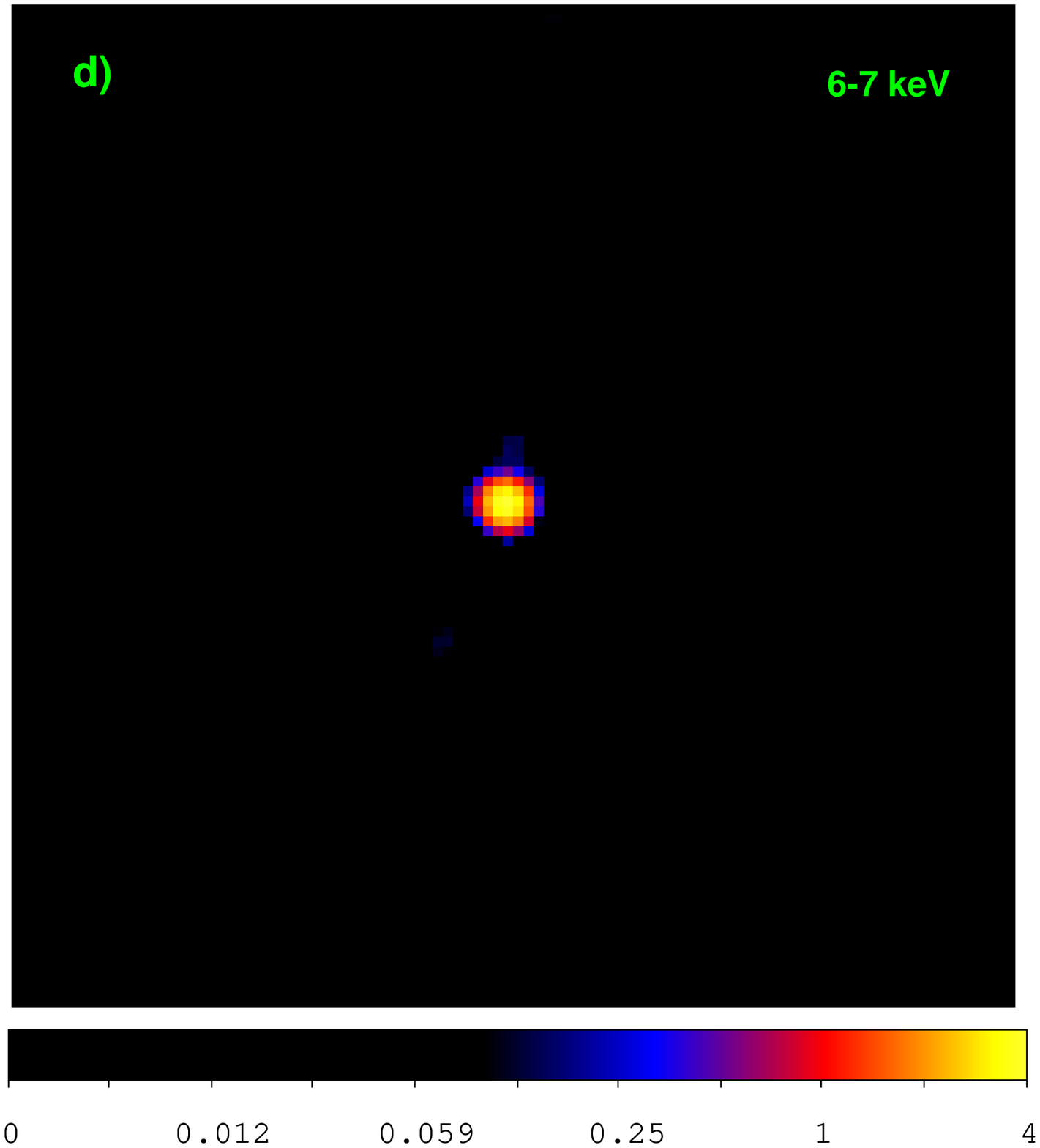,width=0.4\linewidth,angle=0,clip=} \\
\end{tabular}
\caption{The \ch\ images of IRAS 19245--7245 in the energy bands of a) 0.3-2 keV; b) 2-4 keV; c) 4-8 keV; and d) 6-7 keV. North is up, and east is left. These images are merged from the three separated exposures, and smoothed using a Gaussian function with kernel radius of 3 pixels. The annuli of D1 (3-10\arcsec) and D2 (10-20\arcsec) are indicated by the green circles, and the dashed white circle indicates the extraction region (15\arcsec) in \xmm\ observation \citep{2003A&A...398..107B}. The southern (S) and northern (N) nuclei are marked in panel b), as well as one off-nuclear point source (X1). \label{f:merged}}
\end{figure}

In Figure \ref{f:merged}, we show four images of IRAS 19245--7245 by merging the three observations, which are within the energy bands of a) 0.3-2 keV; b) 2-4 keV; c) 4-8 keV and d) 6-7 keV. The images are smoothed (in \texttt{ds9}) with a Gaussian kernel of 3 pixels (1.5\arcsec\ in radius, where 1\arcsec\ corresponds to 1.2 kpc at the redshift of IRAS 19245--7245). We marked the southern (S), northern (N) nucleus and an off-nuclear point source (X1) in panel (b). As shown in panel a) in Figure \ref{f:merged}, there is extended soft X-ray emission contributed by the circumnuclear star formation regions, and we divided the diffuse emission into two regions in shape of annuli with their origin at the southern nucleus: D1 (3-10\arcsec, i.e., 3.5-11.8 kpc) and D2 (10-20\arcsec, i.e., 11.8-23.6 kpc).In the following subsections, we will analyze the imaging and spectra of the nuclei and the diffuse emission. 

Throughout the work in this paper, all \ch\ spectra as well as the associated response (RMF) and ancillary (ARF) files were extracted by running the script \texttt{specextract}, which has slightly different input parameters for point-like source and extended emission. We adopt cosmological parameters of $H_0=70~{\rm km~s^{-1}~Mpc^{-1}}$, $\Omega_{\rm M}=0.27$ and $\Omega_{\Lambda}=0.73$.

\section{Analysis \label{results}}

\subsection{The entire merging system}

IRAS 19254--7245 was observed by \xmmn\ in March 2001 \citep{2003A&A...398..107B} and \suzaku\ in November 2006 \citep{2009A&A...504...53B}. The published X-ray fluxes and spectral properties were derived from spectral analysis of the global emission (here `global' means the entire region including the two nuclei and the extended starburst region). We thus first analyzed the global spectra of our \ch\ observation, and compared them with the \xmm\ and \suzaku\ observations to see if there is any variability. The radius of the extraction region is 15\arcsec, centered on the southern nucleus, which follows the work of the \xmm\ observation. The source region in the \suzaku\ paper has a radius of $\sim 2.9$\arcmin, which covers nearly the entire S3 chip. There are several visually-detected point sources which are located 1-3\arcmin\ away from the southern nucleus of IRAS 19254--7245. Therefore, we did not use this source radius to extract the global spectra. The background emission was chosen from a close-by source-free circular region.

The effective exposure time after flare filtering is shown in Table \ref{t:summary}. Photons with energy greater than 8 keV were ignored in the spectral fitting, since the effective area of the ACIS chips is very small above 8 keV. We obtained 1850 photons in total for all three exposures. The spectrum of each observation was grouped to at least 10 counts per bin to allow the use of $\chi^2$ statistics.  

Following the work of \cite{2003A&A...398..107B,2009A&A...504...53B}, we fitted each spectrum using a model of thermal plus power-law with an additional Fe K$\alpha$ line. The weighted average Galactic {\hbox{H\ {\sc i}}} column density is $N_{\rm H,G}=5.95\times 10^{20}~\rm{cm}^{-2}$ \citep{1990ARA&A..28..215D,2005A&A...440..775K}, and is set to be the lower boundary of the obscuration during the fit. We fixed abundance in the thermal component to solar. The photon index of the power-law continuum was determined by ignoring photons with energy greater than 6 keV and with no Gaussian component added in the model. We then fixed the photon index and used the photons in energy band 6-8 keV to fit the iron emission line. 

\begin{figure}
\centering
\epsfig{file=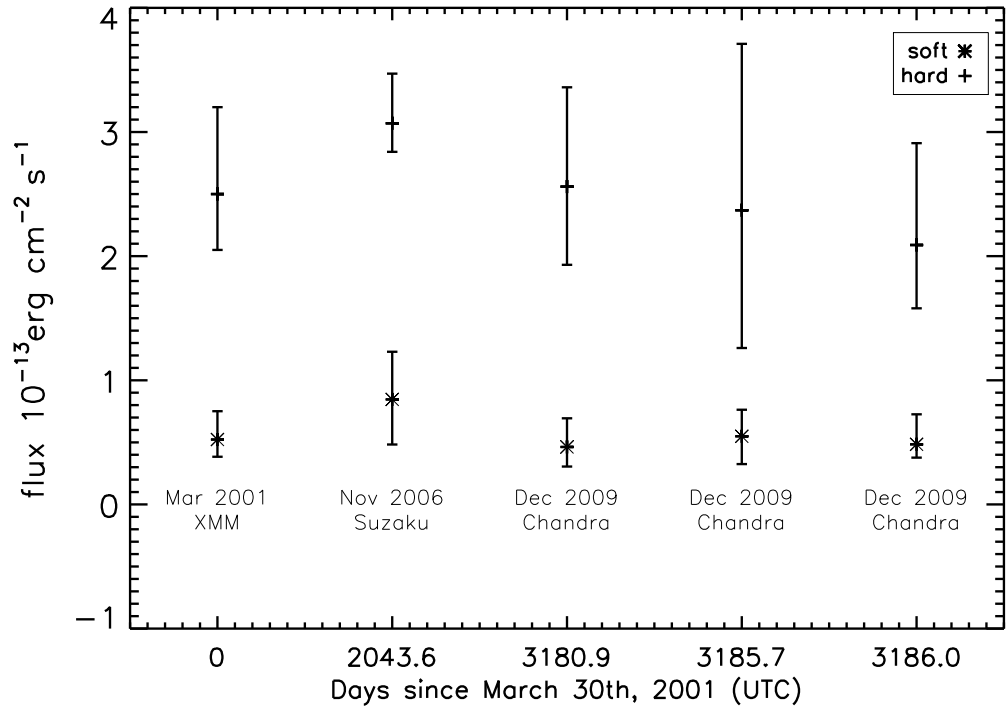,width=0.8\linewidth}
\caption{The soft ({\it asterisk}) and hard X-ray ({\it plus}) flux of each observation.  The points correspond in chronological order to \xmm, \suzaku\ and \ch\ observations. X-axis indicates the days since the date of \xmm\ observation, and is not in linear scale. \label{f:var}}
\end{figure}

We plot the 0.5-2 keV and  2-10 keV fluxes of the \ch\ and the other two observations by \xmm\ and \suzaku\ in Figure \ref{f:var}. The observed X-ray luminosities reported in \xmm\ observation \citep{2003A&A...398..107B} were about a factor of two higher because of the cosmological parameters used in their paper, so we re-derived the luminosities by using the latest parameter values above. 

Since there is no significant variability of either the soft or hard X-ray flux between each exposure of the \ch\ observation, it is possible for us to fit their combined spectra. The spectra were summed using {\tt addspec} in {\tt HEASoft FTOOLS}. Again, the combined spectra were grouped for at least 10 counts per bin, and $\chi^2$ statistics was used in fitting. The best-fitting parameters are shown in Table \ref{t:fit}, as well as the observed soft and hard X-ray luminosities contributed by each component, where errors are determined at the 90\% confidence level for one parameter of interest.

In order to put more constraints on the spectral properties of IRAS 19254--7245, we add the \xmm\ and \suzaku\ data to the spectral fitting. We loaded the PN spectrum of \xmm\ (as the first group), the XIS-FI and XIS-BI spectra of \suzaku\ (as the second group), and the three ACIS-S spectra of \ch\ (as the third group) to perform a spectral fit of the time-averaged spectra. The parameters of the thermal and power-law components are tied together in different spectrum groups except for the relative normalization via a constant multiplicative factor. The temperature and power-law index parameters are close to the results of \ch\ data only but with smaller uncertainties, as shown in Table \ref{t:fit}. The luminosities are calculated as the mean of the three observations. The plasma abundance was initially fixed at solar, and we then allowed it to vary as a free parameter in this fit. The best-fitting abundance was $Z=0.34 Z_{\odot}$, and was poorly constrained. We thus fix this parameter to be at solar in this and the following spectral analysis, since the model fitting is not very sensitive to it. We first tied the normalization of the iron line component in all three observations, and the line energy was found to be at 6.68 keV, consistent with the \ch\ only fit above. However, the line EW was only $\sim 0.78$ keV compared to 1.71 keV we obtained above, which is a hint for variability of the iron line emission. Thus, we allowed the normalization of the iron line component to be free between the three groups. As the EW and luminosity of iron line differ between three observations, we leave them as blank in Table \ref{t:fit}, and we will discuss the variability in the following sections.
 
\subsection{Southern nucleus}

The southern nucleus is known to host a Compton-thick AGN by previous X-ray observations \citep{2000MNRAS.314..589P,2003A&A...398..107B,2009A&A...504...53B}. Those X-ray satellites have low spatial resolution, and the detected X-ray emission from the AGN was contaminated by the host galaxy. We are now able to disentangle the AGN from its nuclear starburst region in our \ch\ observation. We extracted the spectra of the AGN in the southern nucleus from a circular cell with radius 2.5\arcsec. The nucleus is a point-like source, so an energy-dependent point-source aperture correction was applied to the source ARF file. A local background spectrum was extracted from an annulus centered on the nucleus with an inner radius of 3\arcsec\ and an outer radius of 4\arcsec. 

We did a simultaneous joint fit to all three spectra with a model of an absorbed power-law plus a Gaussian at first.  Then, an additional thermal component ({\tt APEC}) was added to the fit. The abundance was again fixed at the solar value, and we found a best-fit temperature of the plasma $kT=0.79_{-0.23}^{+0.39}$ keV. $\chi^2$ decreases from 90.0 for 105 d.o.f (without the thermal component) to 81.7 for 103 d.o.f. Compared to the power-law plus Gaussian model, this model was preferred at 99.3\% confidence level according to the $F$-test. 

\begin{figure}
\centering
\epsfig{file=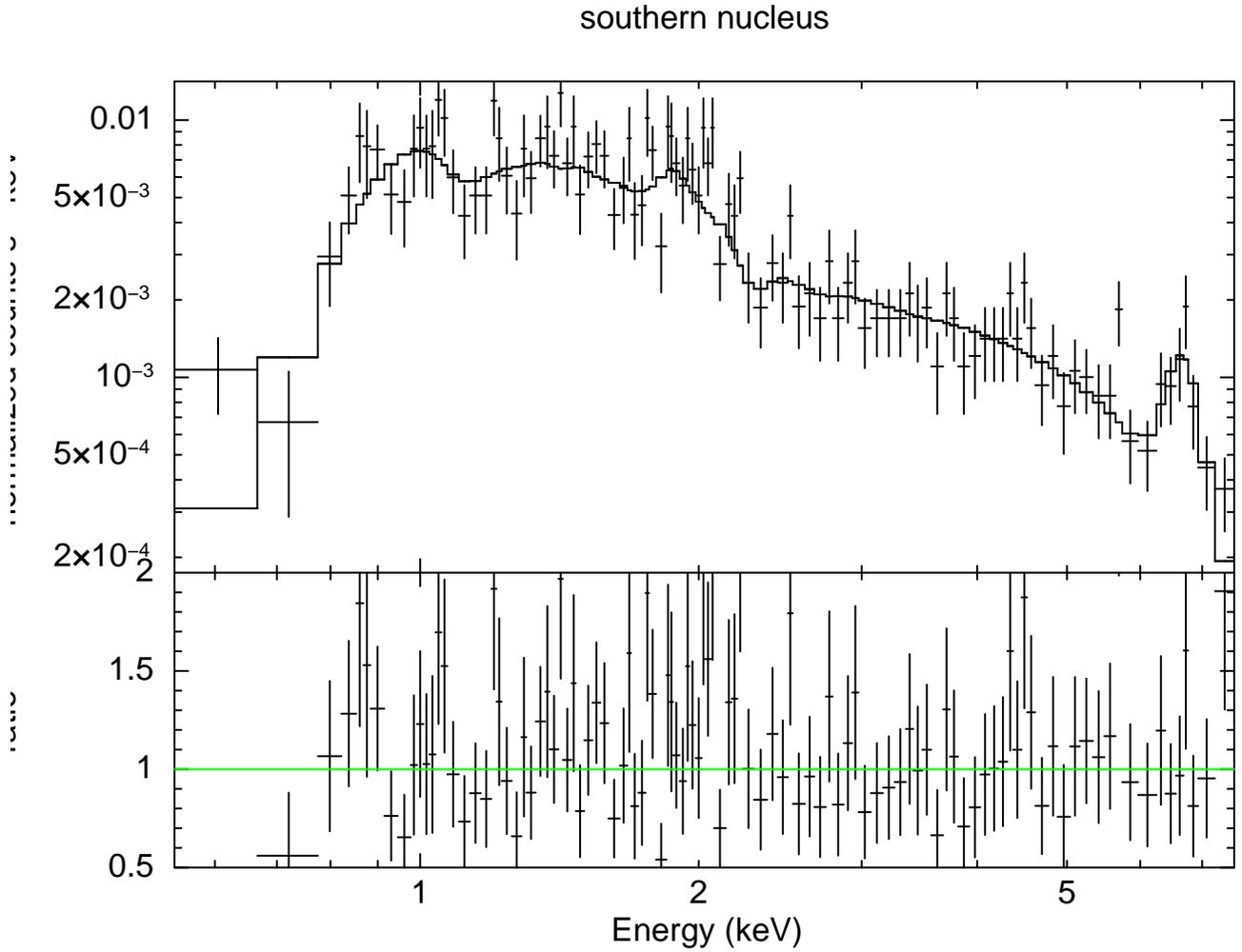,width=0.8\linewidth,angle=-90,clip=} 
\caption{\ch\ spectrum of the southern nucleus. The spectral model consists of a thermal, a power-law and an emission line component. Photon energy is de-redshifted to the rest frame. {\it Top,} data and the model; {\it bottom,} ratio of data to model.\label{f:agn}}
\end{figure}

We also fit the summed spectrum and the spectral parameters are close to the joint fit above.  We list the best-fitting parameters of both the joint fit and the summed-spectrum fit in Table \ref{t:fit}.  The plot of the summed spectrum is shown in Figure \ref{f:agn}. The flat continuum ($\Gamma \sim 1.2$) of the hard X-ray spectrum and strong iron emission line at 6.7 keV with EW $\sim$ 1.5 keV are consistent with the previous X-ray observations.

An interesting question is if the soft flux in this spectrum is due to circumnuclear starburst emission or is more directly associated with the AGN. To test how this emission scales with extraction region size we also extracted the spectrum of southern nucleus in a smaller region with radius of 1.5\arcsec. However, the background was extracted from the larger source region (with radius = 2.5\arcsec, the same region as the source extraction region above), scaled to the area of the 1.5\arcsec\ circle. The rationale is that any component that is extended over the 2.5" region with an $\sim$ constant surface brightness would subtract out completely, while nuclear point-like emission by reduced only by the ratio of the areas of the two regions, i.e., $1.5^2/2.5^2 \sim 1/3$.  The temperature of the thermal component is $kT=0.78_{-0.48}^{+0.40}$, nearly the same as the value above but with larger uncertainty. The power-law normalization is two thirds of that in the spectral fit of the larger extraction region case, and so is the iron line normalization, consistent with both of these components being point-like. The ratio for the {\tt APEC} normalization holds the same scaling relationship as that of the power-law normalization in the two cases. This indicates that either there is also point like, i.e., dominated by a thermal source in the innermost part of the nucleus or the AGN emission is mixed with the ambient star-formation and we cannot resolve it with \ch. If the latter is true, we would expect consistency with the diffuse emission from the extended starburst region, which we will analyze in section \ref{sb} below.

\subsection{Northern nucleus}

Although the northern nucleus is relatively weak in X-ray emission compared with the southern one and is believed not to harbor an AGN, it is still interesting to investigate its X-ray properties. We positioned the northern nucleus by visual examination and extracted the spectra within a circle of radii of 1.5\arcsec\ for Obs ID 10538 and 11797. It could not be visually detected in the observation of Obs ID 11798, and the spectrum was extracted from the region used in Obs ID 10538, which has the brightest northern nucleus in three exposures. 30 photons in total are collected in 0.3-8 keV range. We summed the spectra and fitted the combined spectrum with an absorbed power-law model using {\it Cash} statistic \citep{1979ApJ...228..939C}. Although {\it Cash} statistics is designed for unbinned data, we grouped the spectra to at least 1 count per bin, which could improve the performance of the fit \citep{2005ApJ...633..664T}. The photon index is found to be $\Gamma=1.2_{-1.1}^{+1.5}$. The observed 2-10 keV luminosity derived from this model is $L_{\rm 2-10keV}=5.0\times 10^{40}~{\rm erg~s}^{-1}$. The spectral parameters are listed in Table \ref{t:fit}. 

\subsection{Diffuse emission\label{sb}}

\subsubsection{Imaging}

\begin{figure}
\centering
\begin{tabular}{cc}
\epsfig{file=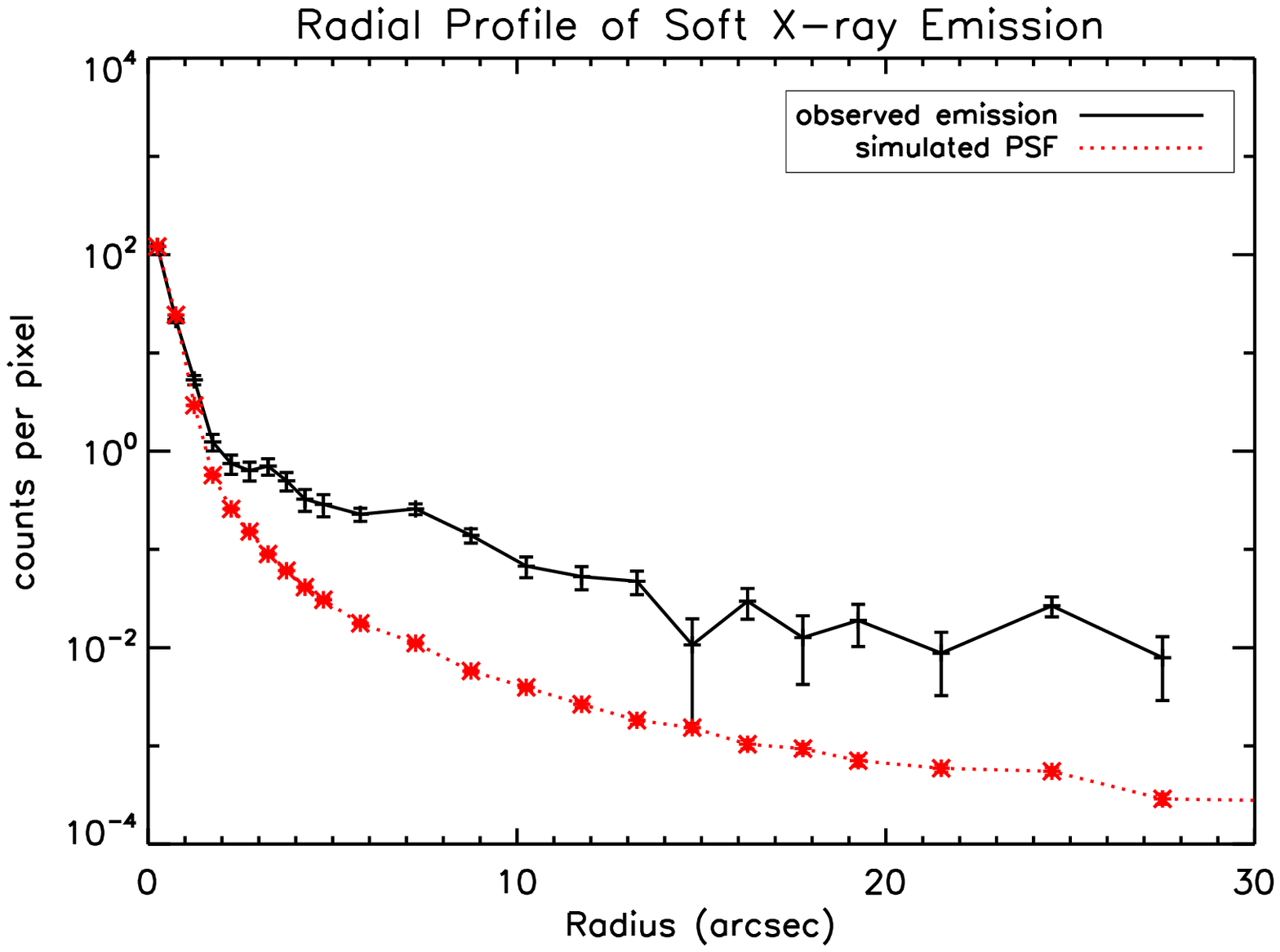,width=0.4\linewidth,angle=0,clip=} &
\epsfig{file=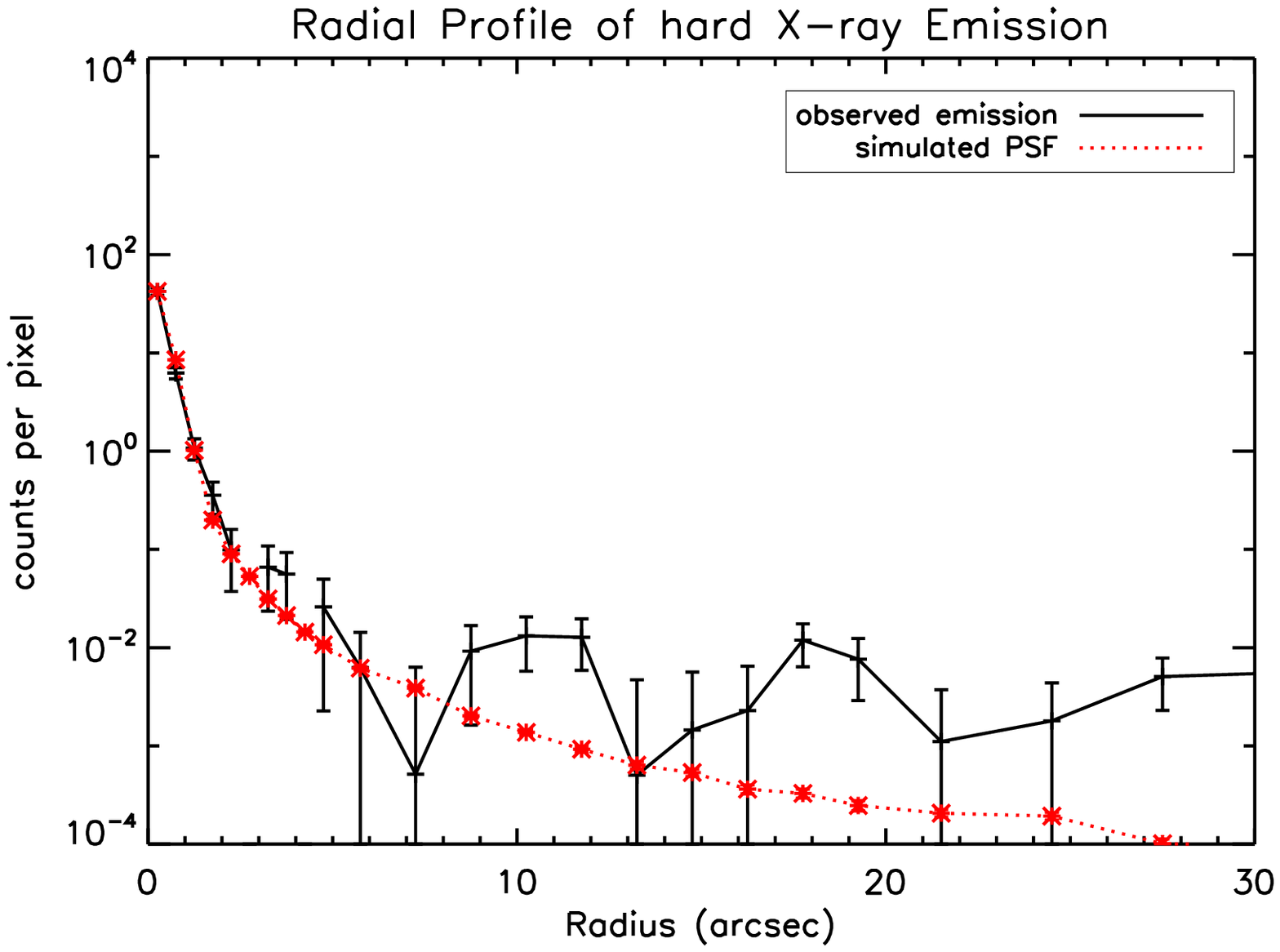,width=0.4\linewidth,angle=0,clip=} \\
\end{tabular}
\caption{Radial count distributions of the observed extended structure from the centroid of the southern nucleus (in black) and the simulated PSF (in red) in both soft (left) and hard (right) X-ray bands.
\label{f:profile}}
\end{figure}

Thanks to the high spatial resolution of \ch\, the diffuse emission from the star formation regions could be imaged (see Figure \ref{f:merged}). In order to study the extended structure, we investigated the brightness distribution of IRAS 19254--7245. We first simulated an observation of a point source with the same spectral properties as the merged event using the {\it Chandra Ray Tracer} \citep[ChaRT;][]{2003ASPC..295..477C} to trace the rays through the \ch\ X-ray optics, and created an image of the PSF using MARX (version 4.5). The comparison between the observed radial profile of the extended structure and the simulated PSF in both soft and hard X-ray bands is shown in Figure \ref{f:profile}, where the X-ray centroid of the southern nucleus was used as the origin. In the left panel of Figure \ref{f:profile}, we see that the observed distribution of the soft X-ray emission is rather close to the simulated one within the radius of $\sim$2\arcsec, and for radii between 3\arcsec and 20\arcsec, there is a significant excess of the observed counts with respect to the simulated PSF. In the hard X-ray band shown in the right panel of Figure \ref{f:profile}, the radial profile of the extended emission is consistent with the simulated PSF.

\subsubsection{Spectral analysis}

In order to extract the spectrum of the circumnuclear starburst, the emission from the AGN was excised within a circle of 3\arcsec\ radius, and other point sources detected by {\tt wavdetect} were also excluded. We extracted the spectra of the star-forming region in two annuli centered on the southern nucleus: one has the inner radius of 3\arcsec\ and outer radius of 10\arcsec\ (D1); the other one has inner radius of 10\arcsec\ and outer radius of 20\arcsec\ (D2), as described in Section \ref{data}. The background spectra were extracted in a concentric annulus with inner radius of 40\arcsec\ and outer radius of 50\arcsec. The spectra of three exposures were then added to improve the signal-to-noise ratio. 

{\it Soft emission:} The extended star-forming region is dominated by soft emission, as shown by the comparison between the soft and hard X-ray images in Figure \ref{f:merged}. We first ignored the hard X-ray photons with energy greater than 3 keV, and fitted the spectrum of the soft emission. It was initially characterized by a thermal plasma with solar abundance. The resulting temperature was $kT=0.77_{-0.09}^{+0.08}$ keV for D1 and $kT=0.91_{-0.60}^{+0.48}$ keV for D2 (see Table \ref{t:sb}). The plasma temperatures are consistent with the spectral fits of \xmm\ and \suzaku\ data, and are within the the typical range of starburst galaxies \citep[e.g.,][]{1999ApJS..120..179P,2005ApJ...628..187G}. The temperature of the diffuse emission in D1 is also close to the value for the AGN spectra, indicating that the emission from the southern nucleus is probably also exhibiting thermal emission from the circumnuclear star-forming regions. We then investigated the possibility of introducing another thermal component associated with an absorber in the spectral fitting since such a feature is found in other ULIRGs \citep{1998ApJS..118..401D,1998MNRAS.297.1219I,2002ApJ...569..655G}. This second plasma component is colder, with a temperature of $kT=0.34_{-0.06}^{+0.23}$ keV for D1 and $kT=0.27_{-0.12}^{+0.18}$ keV for D2. The best-fitting parameters are listed in Table \ref{t:sb}. The two-temperature model does not improve the spectral fit much, and the normalization of the cooler component has a very large uncertainty, so we thus adopt the results of the single temperature model in this paper.

\begin{figure}
\centering
\begin{tabular}{cc}
\epsfig{file=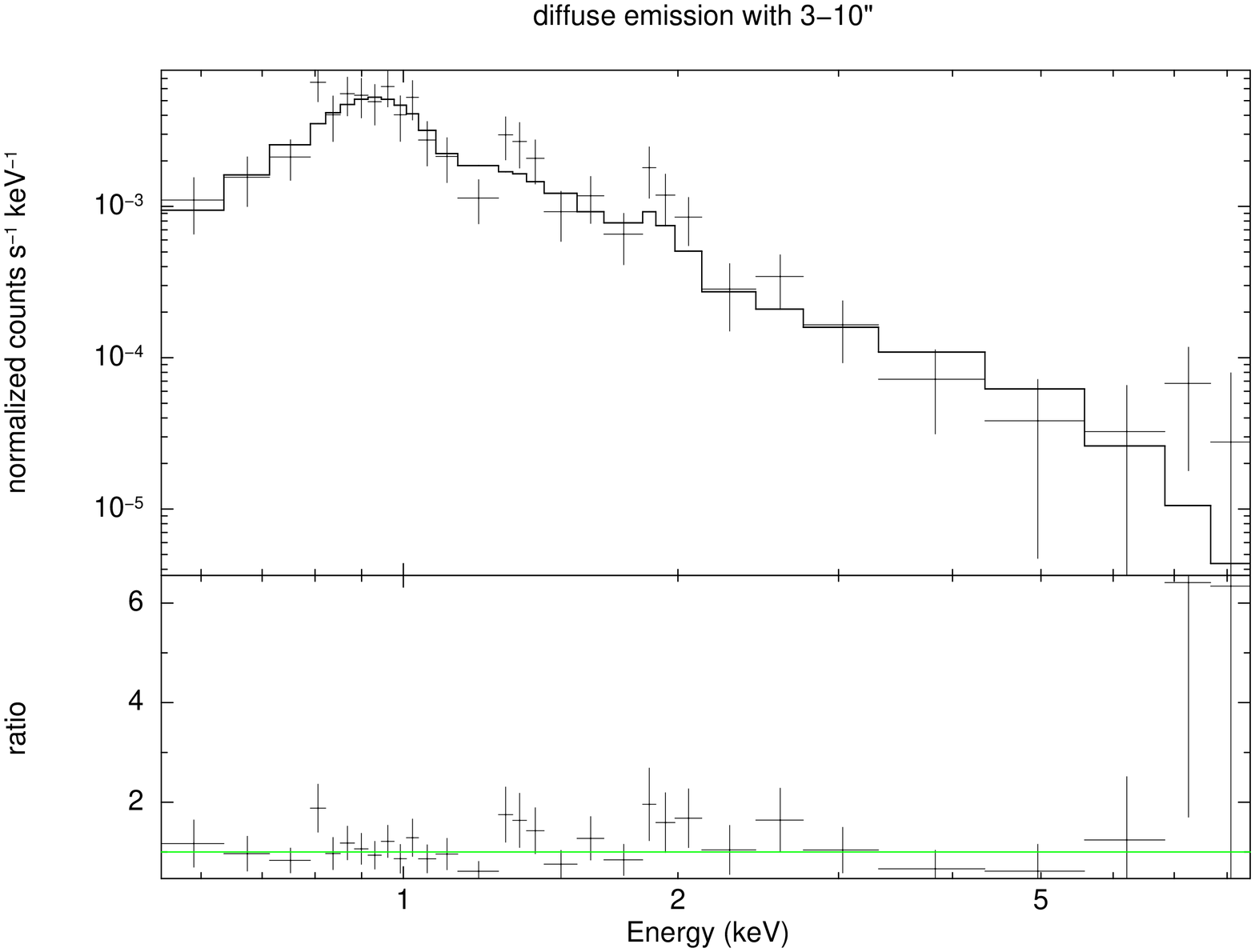,width=0.4\linewidth,angle=-90,clip=} &
\epsfig{file=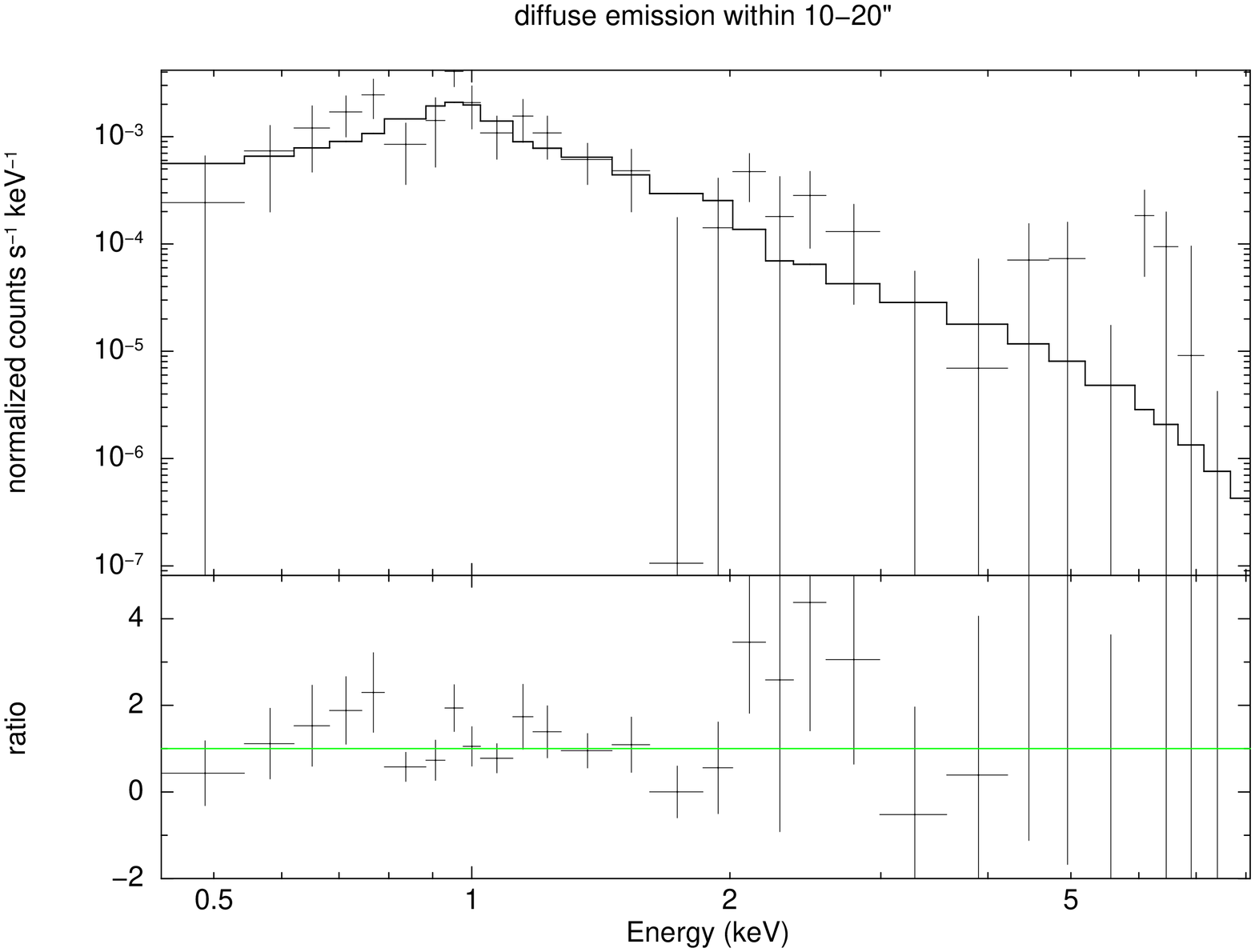,width=0.4\linewidth,angle=-90,clip=} \\
\end{tabular}
\caption{Spectral fits of the diffuse emission in annuli of 3-10\arcsec\ (left) and 10-20\arcsec\ (right), where the southern nucleus is at the center. 
\label{f:sb}}
\end{figure}

{\it Hard emission:} The hard continuum was fitted by a power-law, and we included the soft emission in the spectral fitting because the power-law parameters could be determined much more accurately in this way. For the inner region D1, the photon index is $1.73_{-0.56}^{+0.60}$, a typical value for AGN. For the outer region D2, we initially got a much steeper index of $2.70_{-0.88}^{+0.87}$. However, the larger errors due to low signal-to-noise makes it consistent with the value in the spectral fit of D1. The soft and hard X-ray luminosities are also listed in Table \ref{t:fit}, and the spectral plots of both regions are shown in Figure \ref{f:sb}. As we see in Table \ref{t:fit}, the thermal emission and power-law emission account for similar amounts of the soft X-ray emission. The hard X-ray luminosity is 20 times smaller than that of the southern nucleus. The best-fit column density along the light of sight is negligible in D2 so we set it to be equal to the Galactic column density, and it is slightly larger in D1. Because the hard X-ray emission in the D2 is much weaker than in D1 and has poorer quality, we extracted the spectra from the combined regions of D1 and D2 and re-fitted the hard spectrum. The power-law index was then found to be close to that in D1.

The spectrum of the diffuse emission shows an excess in the 6-7 keV band. Could it be due to ionized iron emission contributed by very hot gas driven by the starburst \citep{1990ApJS...74..833H,2007ApJ...658..258S}? We show in panel d) in Figure \ref{f:merged} the image within the energy range of 6.0-7.0 keV, which covers the neutral and ionized iron emission line energies in observer's frame. This image shows that the region of iron line emission overlaps the southern nucleus and is not extended. However, due to the low signal-to-noise ratio and limited photons in hard X-ray band, we cannot rule out the possibility that the hard continuum emission is due to another much hotter plasma. If we fit the hard X-ray spectrum with a thermal component, the temperature is $kT\sim 6$ keV with very large uncertainty, and the reduced $\chi^2$ is nearly the same as that in the case of power-law.

\subsection{Iron line emission}

The iron emission line in the Southern nucleus has a relatively broad profile ($\sigma \sim 0.3$ keV), and \cite{2009A&A...504...53B} showed the possibility that it is a blend of multiple lines. It is therefore interesting to further investigate this feature. In order not to wash out the possible fine structure of the iron line profile, we fitted the unbinned time-averaged Chandra spectra using the {\it Cash} statistic.

\begin{figure}
\centering
\begin{tabular}{ccc}
\epsfig{file=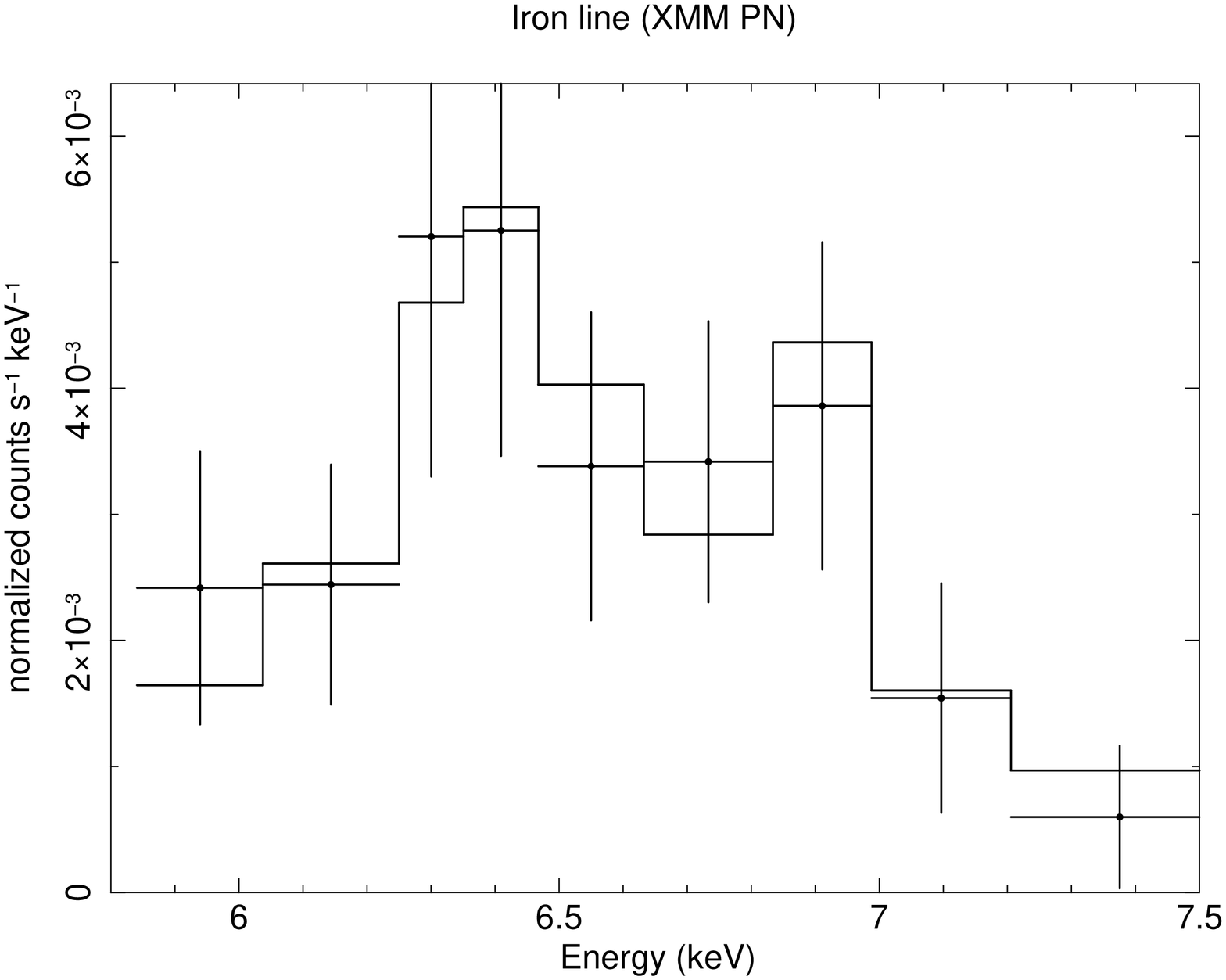,width=0.25\linewidth,angle=-90,clip=} &
\epsfig{file=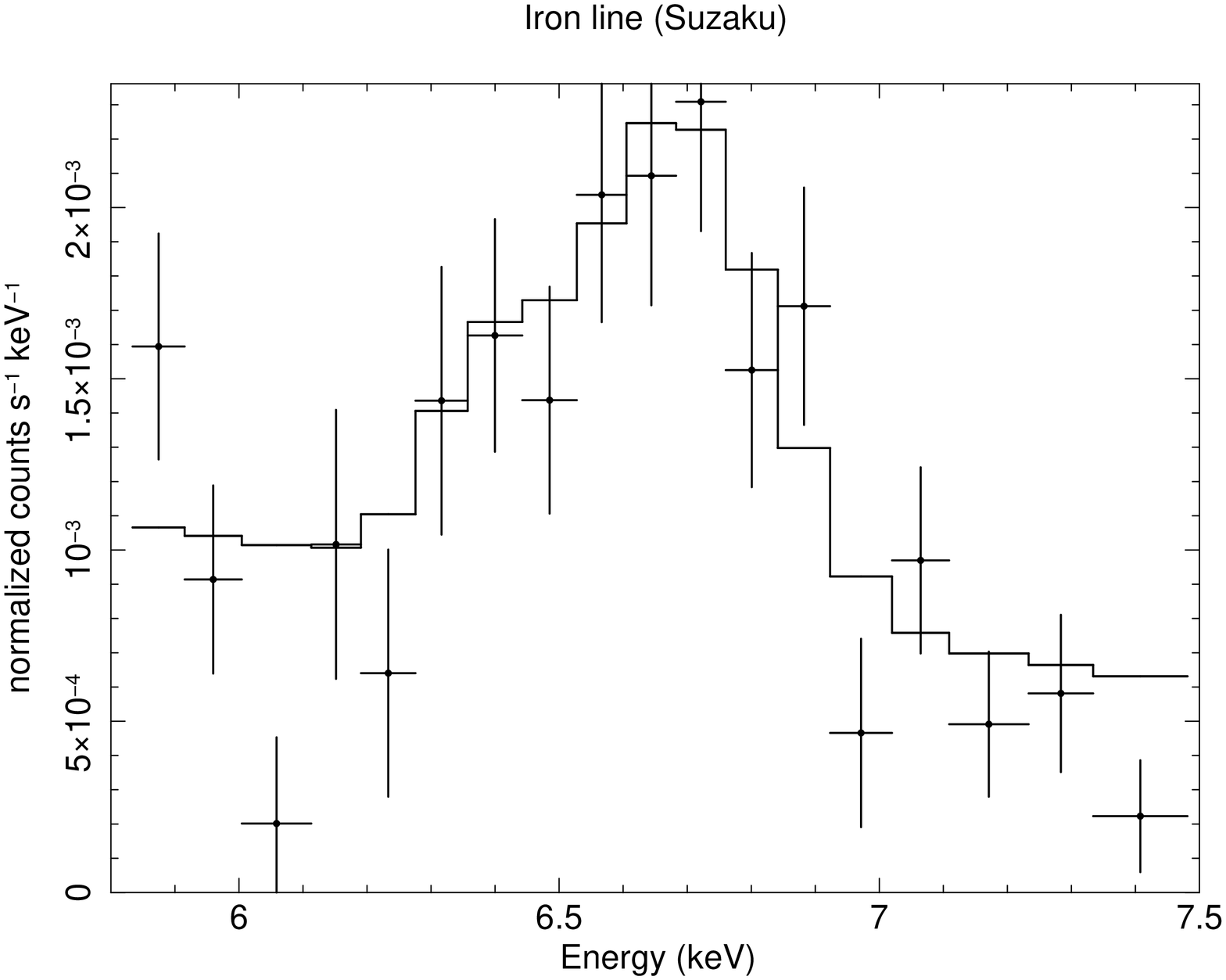,width=0.25\linewidth,angle=-90,clip=} &
\epsfig{file=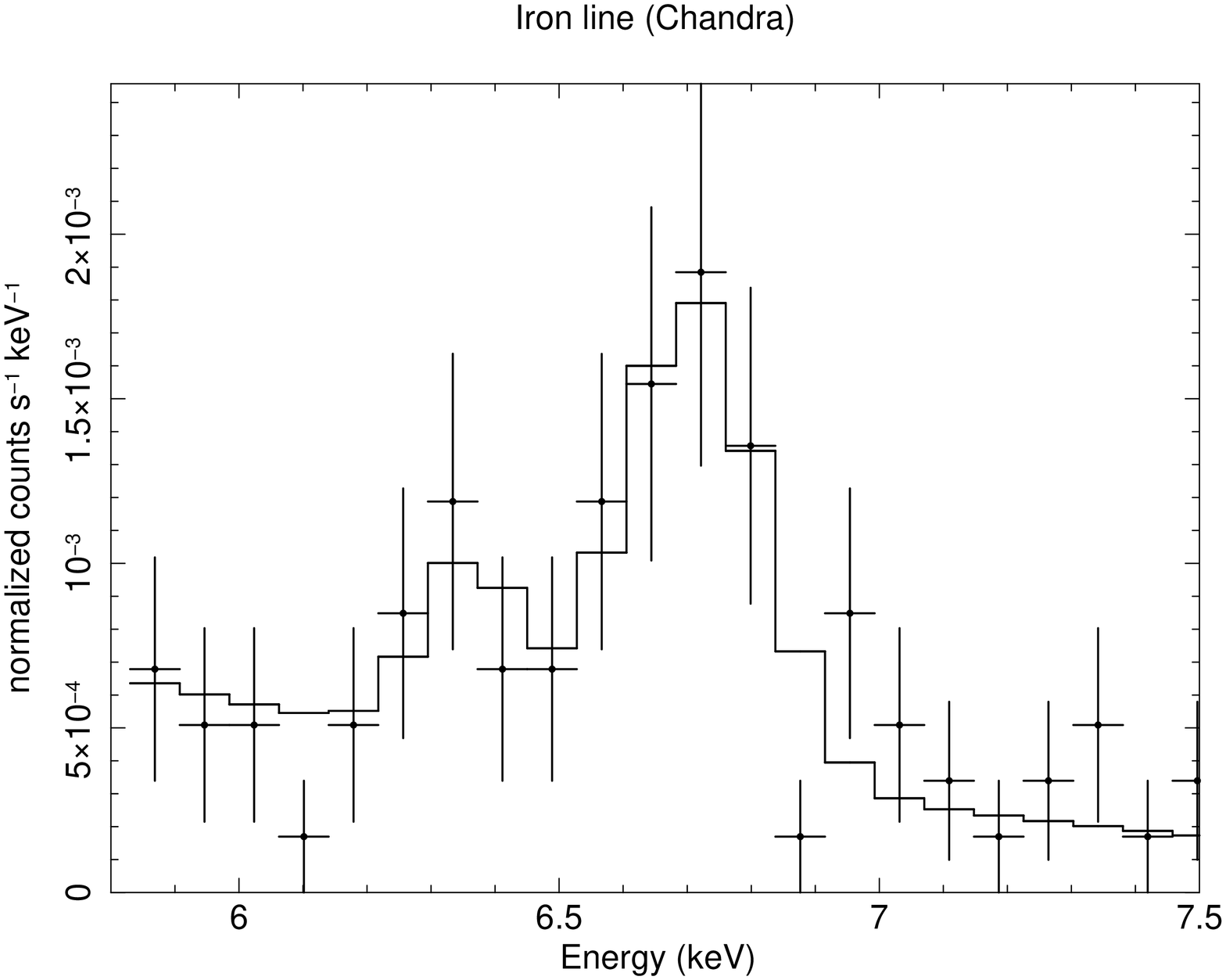,width=0.25\linewidth,angle=-90,clip=} \\
\epsfig{file=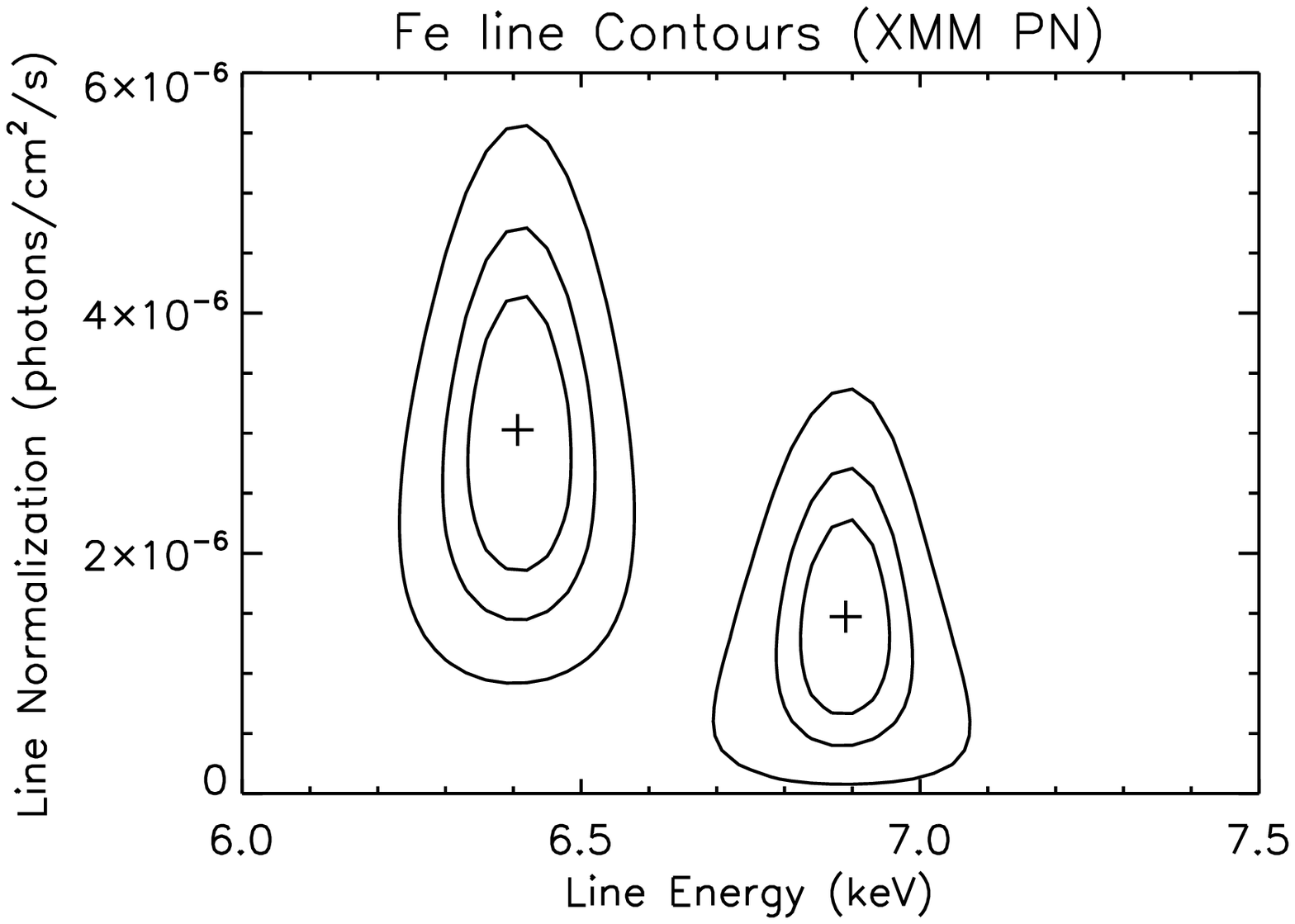,width=0.25\linewidth,clip=} &
\epsfig{file=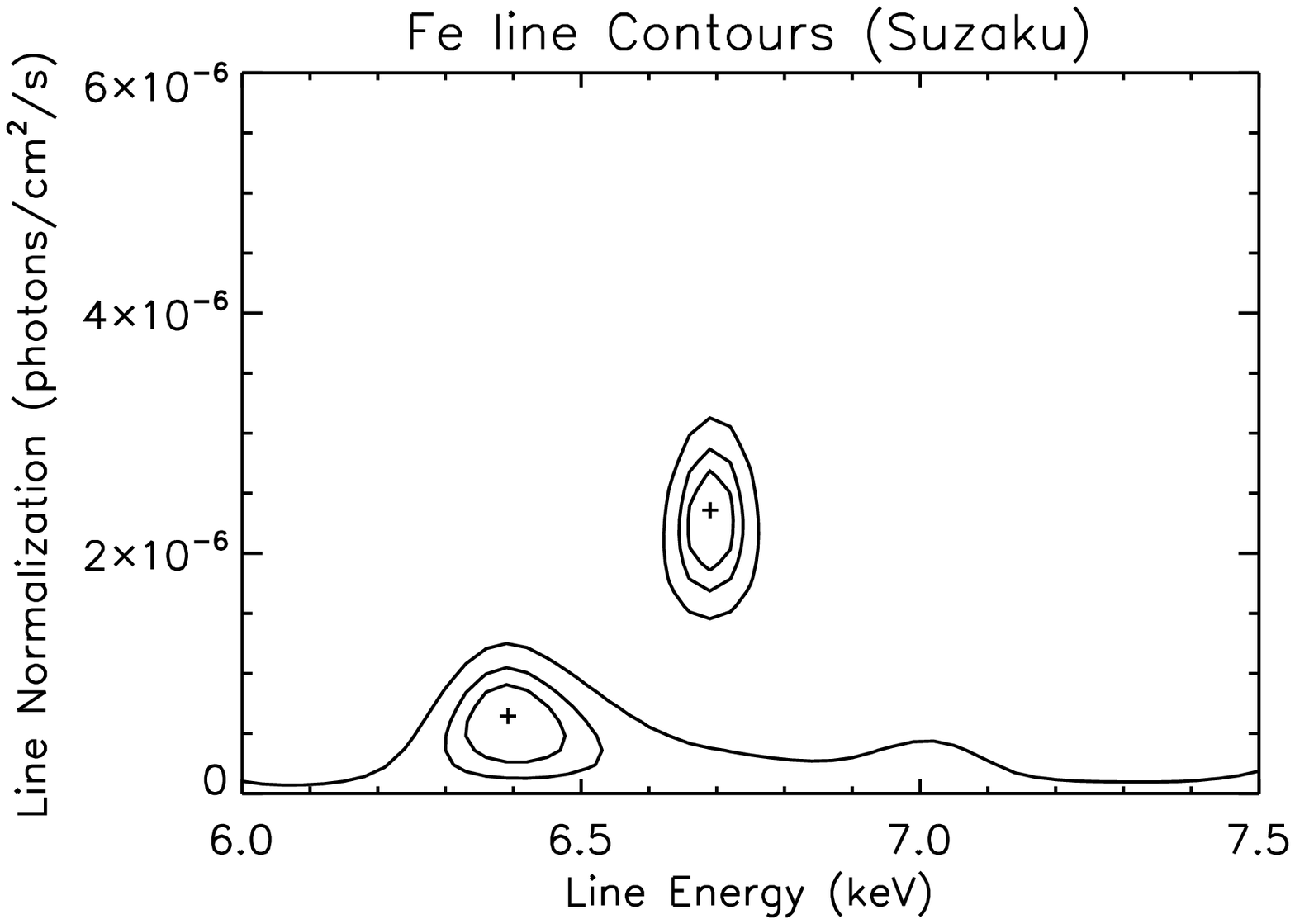,width=0.25\linewidth,clip=} &
\epsfig{file=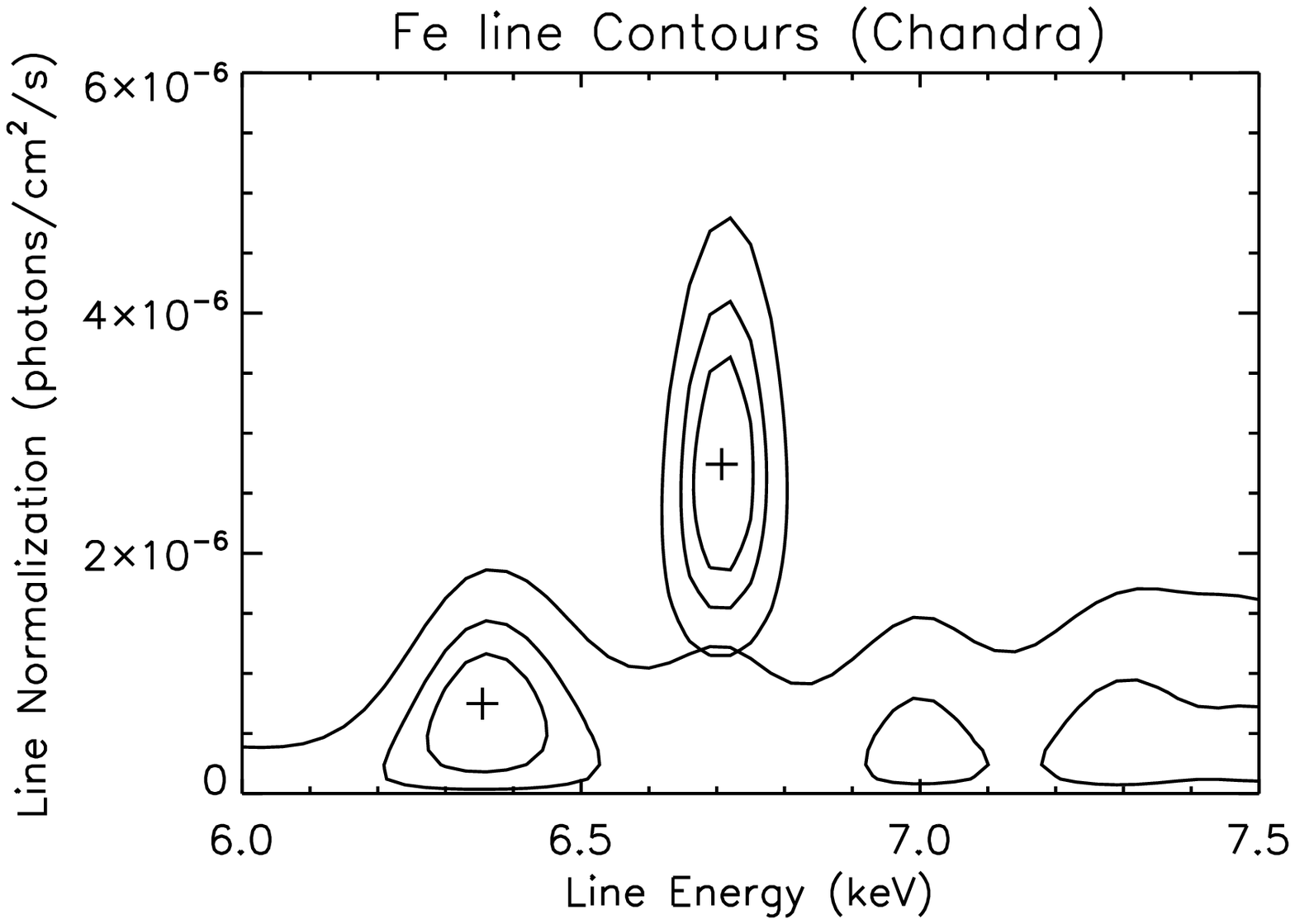,width=0.25\linewidth,clip=} \\
\end{tabular}
\caption{{\it Upper}: Iron emission line profile fitted by double Gaussian lines centered for the \xmm, \suzaku\ and \ch\ observations. The energy is de-redshifted to the rest frame; {\it Lower}: Confidence contour plots for the double Gaussian lines profile. Contours correspond to the 68\%, 90\% and 99\% confidence levels respectively. \label{f:iron}}
\end{figure}

We used a power-law continuum with the photon index fixed at $\Gamma=1.22$ and two gaussian lines to fit the \ch\ spectrum within energy between 3 keV and 8 keV. The centroid energy of two lines are found to be 6.43 and 6.71 keV, and the equivalent widths are 0.17 keV and 0.66 keV respectively, i.e., the neutral iron line is weaker than the ionized iron line. However, this is not the same as when we re-analyzed the \xmm\ data. We fitted the unbinned PN data of the \xmm\ observation in the same way as above. The energy of the ionized iron line is 6.89 keV, which is Hydrogen-like. The EW is 1.16 keV for the neutral line, and 0.44 keV for the ionized line. The double line fit of \suzaku\ data gives similar results as \ch\ for the line energy and EW as the \ch\ observation. We then plotted the confidence contours between the line energy and normalization for both lines at the 1$\sigma$, 2$\sigma$ and 3$\sigma$ levels. The line profile and contours are shown in Figure \ref{f:iron}, where the energy is in the rest frame. We see that the significance of the 6.4 keV line in \suzaku\ and \ch\ observations is only 2$\sigma$, but the lines in the \xmm\ observation could be separately detected at $>99\%$ confidence.

\begin{figure}
\centering
\begin{tabular}{ccc}
\epsfig{file=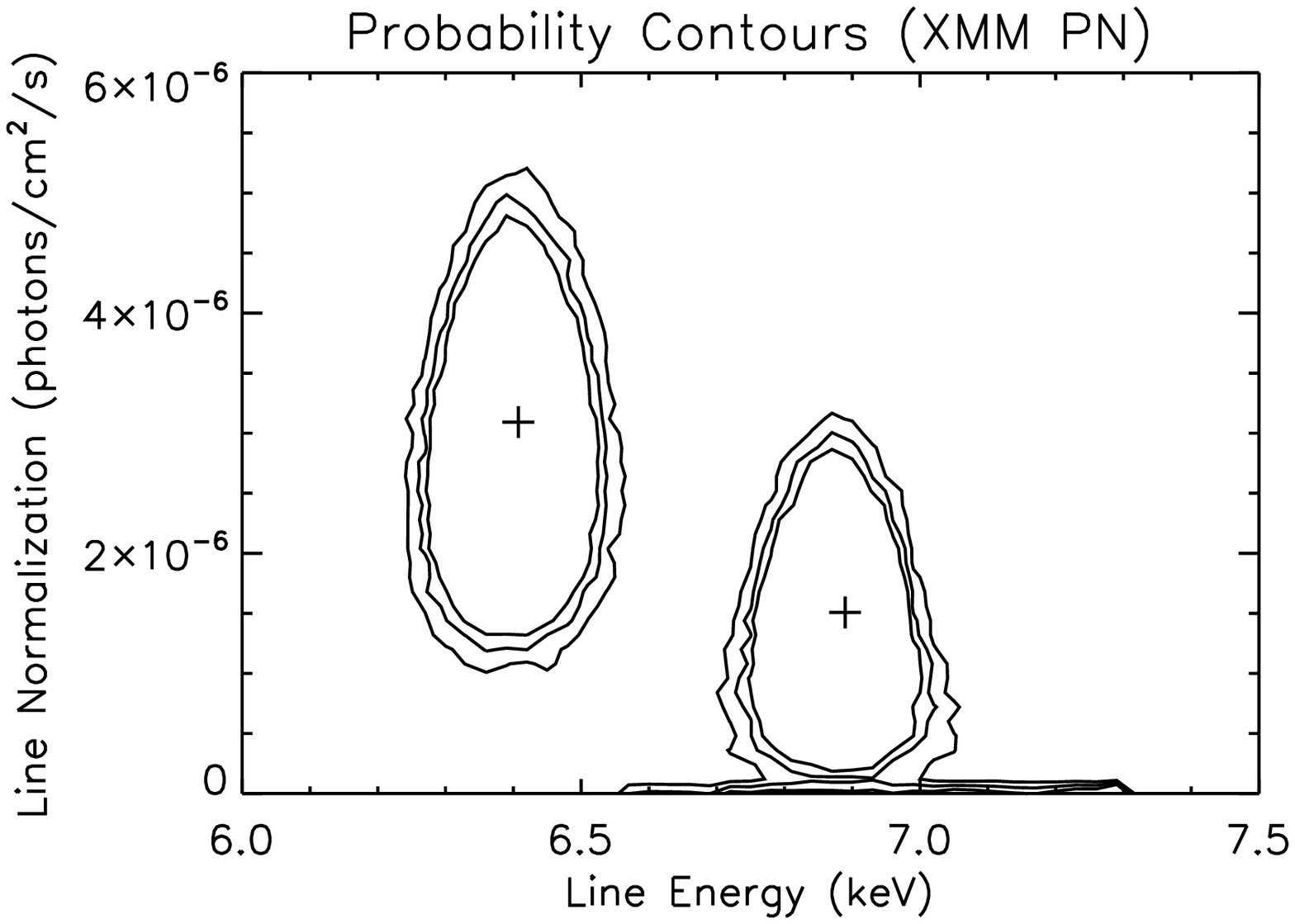,width=0.3\linewidth,clip=} &
\epsfig{file=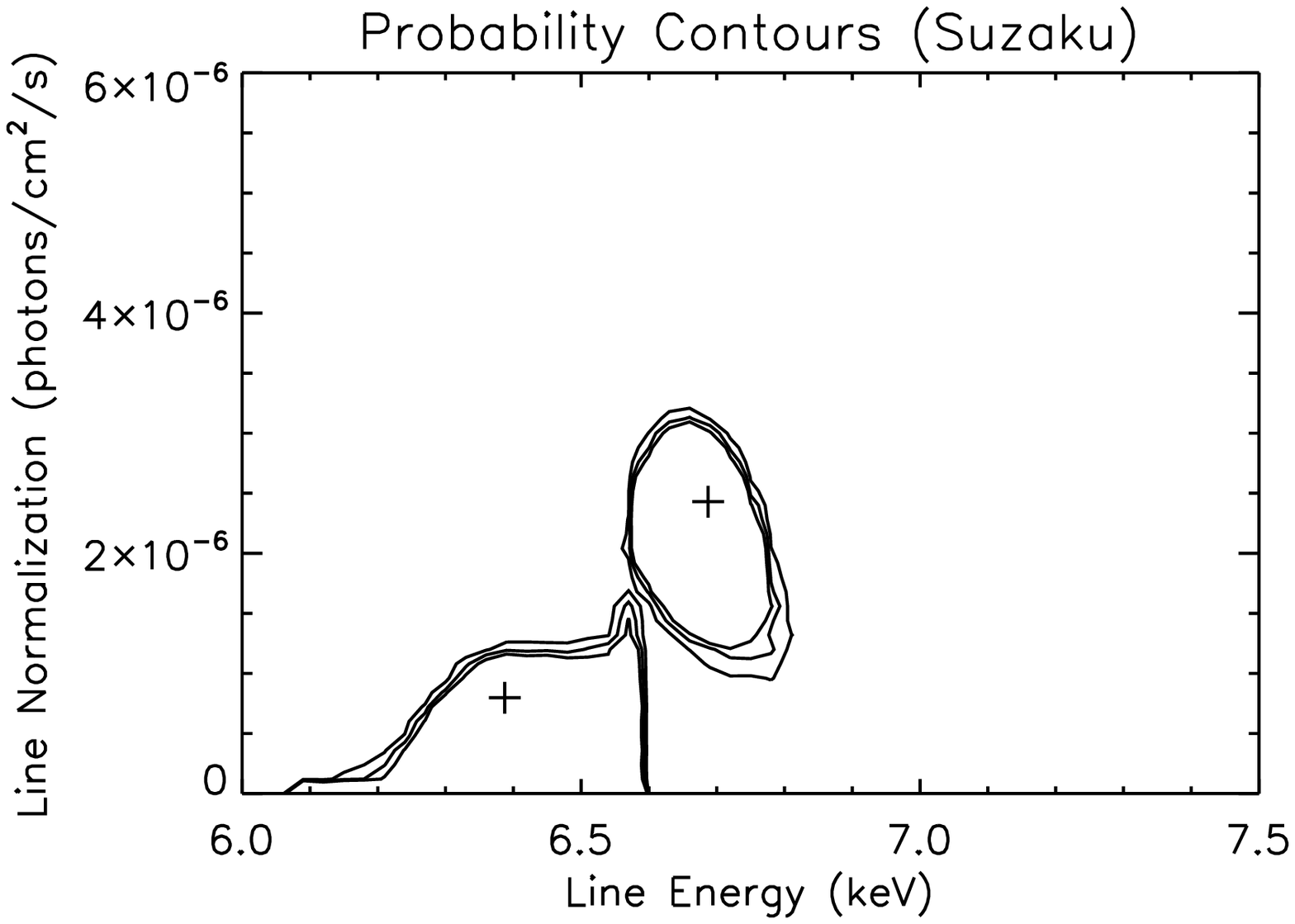,width=0.3\linewidth,clip=} &
\epsfig{file=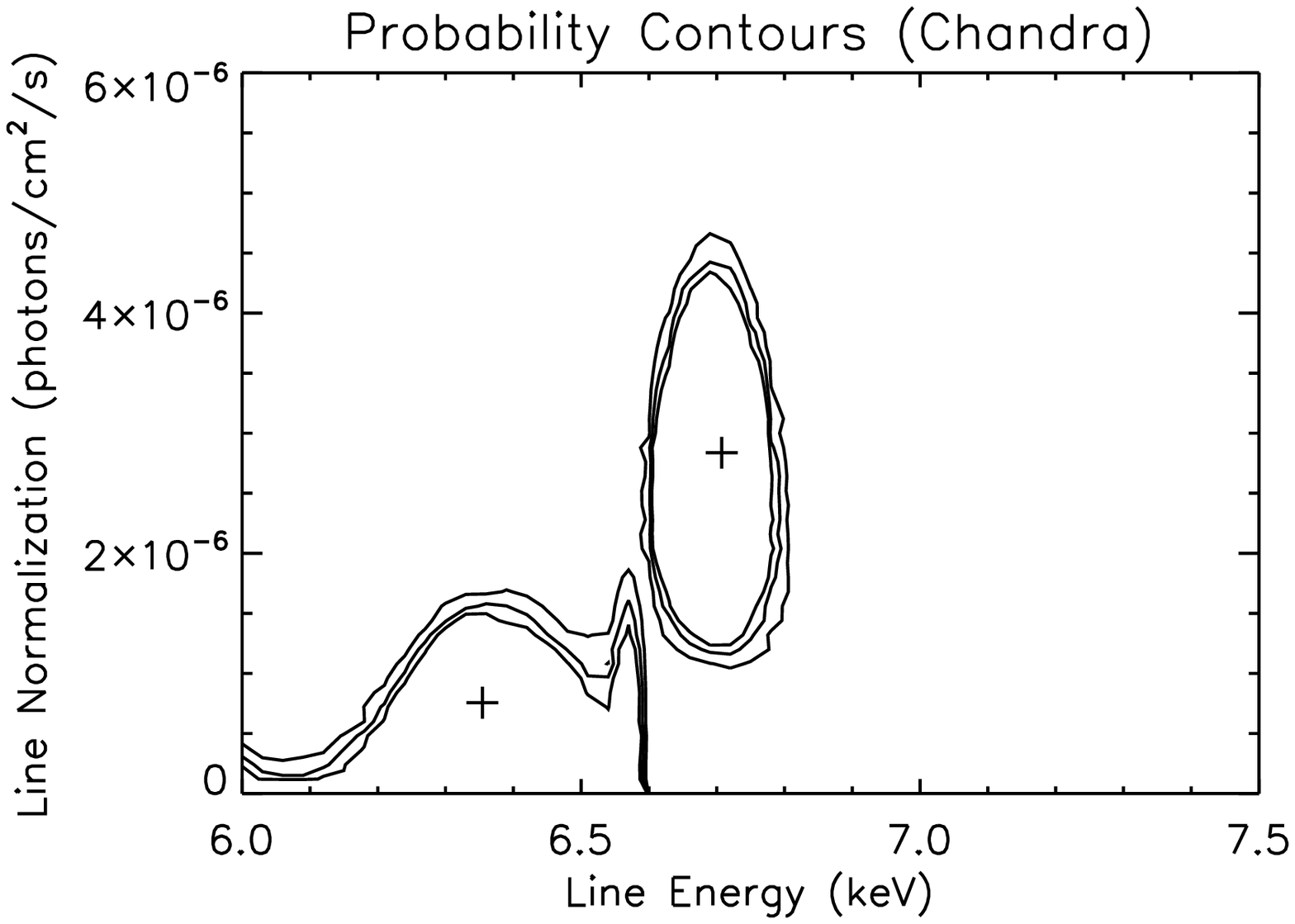,width=0.3\linewidth,clip=} \\
\epsfig{file=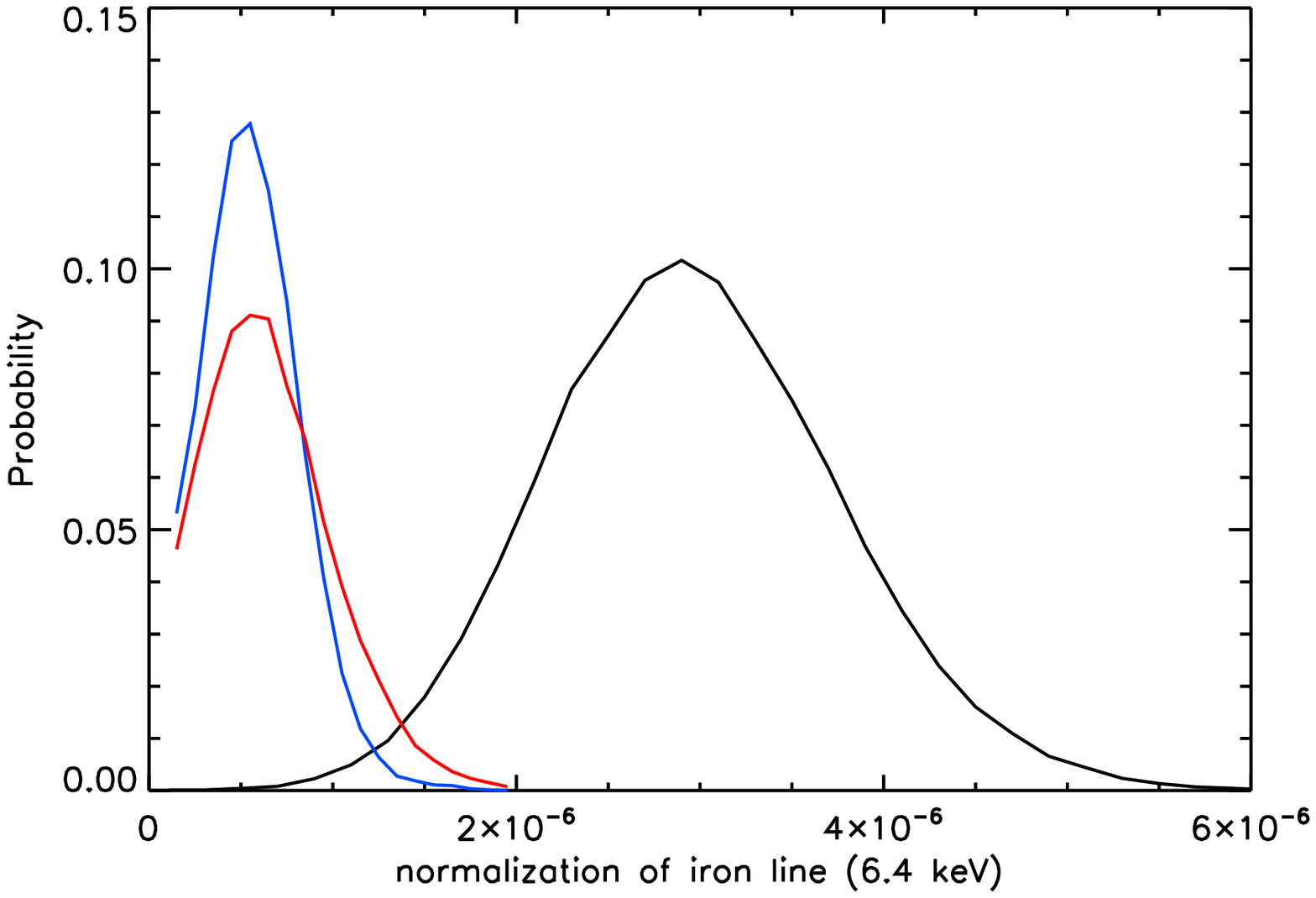,width=0.3\linewidth,clip=} &
\epsfig{file=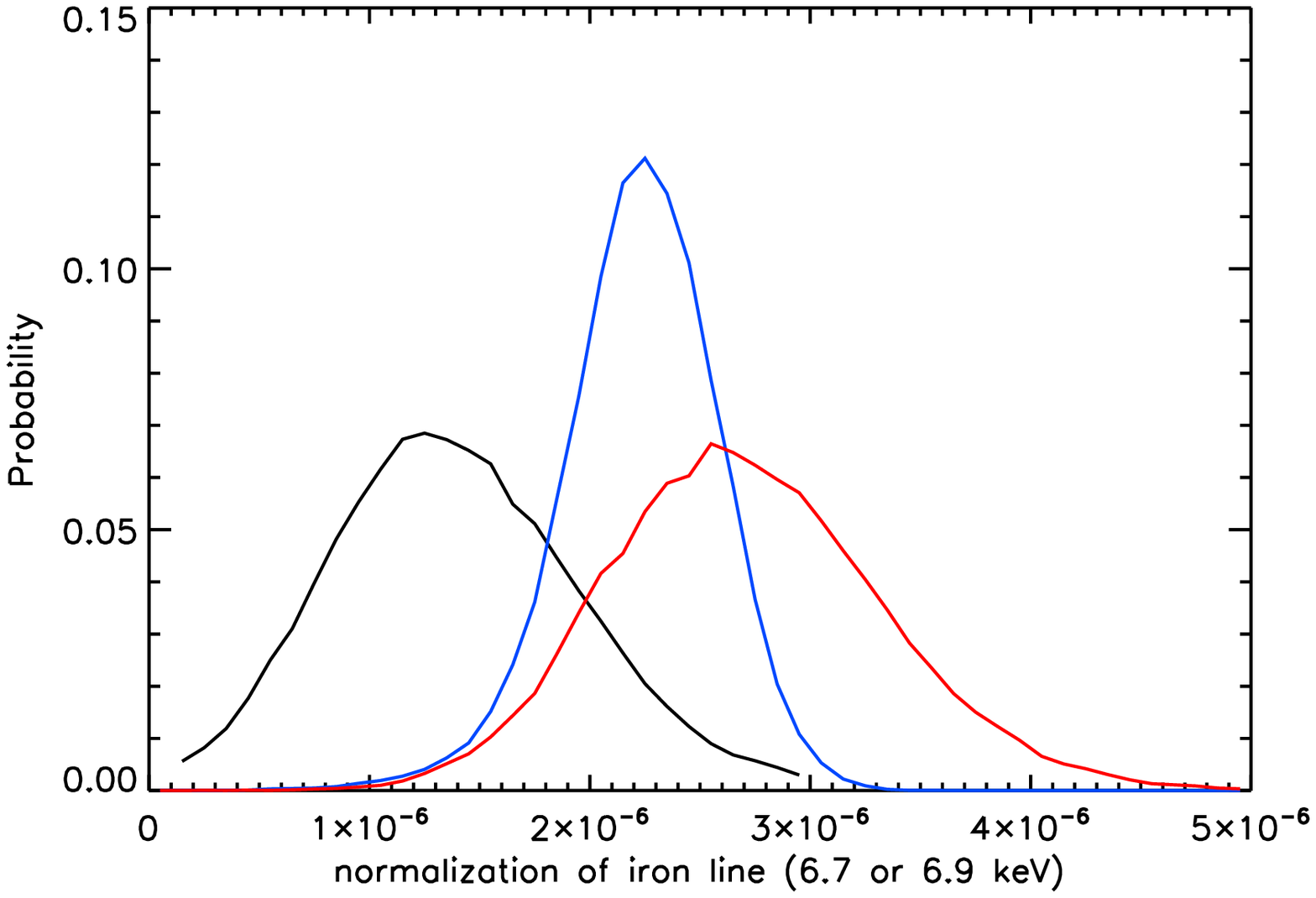,width=0.3\linewidth,clip=} \\
\end{tabular}
\caption{{\it Upper}: Probability contours of the iron lines from MCMC. The contour levels correspond to probabilities of $4\times 10^{-4}$, $8\times 10^{-4}$ and  $1.2\times 10^{-3}$; {\it Lower}: Distribution of the normalization parameter of the iron line component at 6.4 keV (left) and 6.7 or 6.9 keV (right) in \xmm\ (black), \suzaku\ (blue) and \ch\ (red) observations. The parameter probability is in arbitrary units, and the relative scale depends on the bin size.
\label{f:mcmc}}
\end{figure}

We then ran Markov Chain Monte Carlo (MCMC) simulations to test the statistical significance of the lines. Each chain entailed $10^{5}$ iterations, and we ran 5 chains for each spectrum. The parameters and the associated errors are given in Table \ref{t:iron}. The significance of the line was then calculated as the fraction of the iterations where the line normalization parameter was non-zero. We see in Table \ref{t:iron} that the 6.4 keV line in the \xmm\ observation and 6.7 keV lines in the \suzaku\ and \ch\ observations are significant at a level of over 99.99\%. The significance for the 6.9 keV line in the \xmm\ observation is about 97\%, and only less than 90\% for the 6.4 keV lines in the \suzaku\ and \ch\ observations. As in Figure \ref{f:iron}, we plot the probability contours of line energy and line normalization derived from the MCMC analysis in the upper panel of Figure \ref{f:mcmc} at probability levels of $4\times 10^{-4}$, $8\times 10^{-4}$ and $1.2\times 10^{-3}$. Histograms showing the distribution of each line normalization is also shown in the lower panel of Figure \ref{f:mcmc}. The left plot shows the neutral 6.4 keV line normalization distribution and the right plot shows the ionized line (6.7 or 6.9 keV) distributions. Lines in black, blue and red indicate the observation of \xmm, \suzaku\ and \ch, respectively. The peak of distribution of the 6.4 keV line overlaps in the \suzaku\ and \ch\ observation, but is obviously separated from the \xmm\ observation, giving a $\sim 90\%$ probability of the line variability. However, the variability of the ionized iron line is not significant as indicated by the right panel of Figure \ref{f:mcmc}. 

We also tried to use three lines to fit the iron emission spectra. However, the data quality does not allow us to investigate the possibility that both He-like and H-like lines are present.

\subsection{Off-nuclear point source}

\begin{figure}
\centering
\epsfig{file=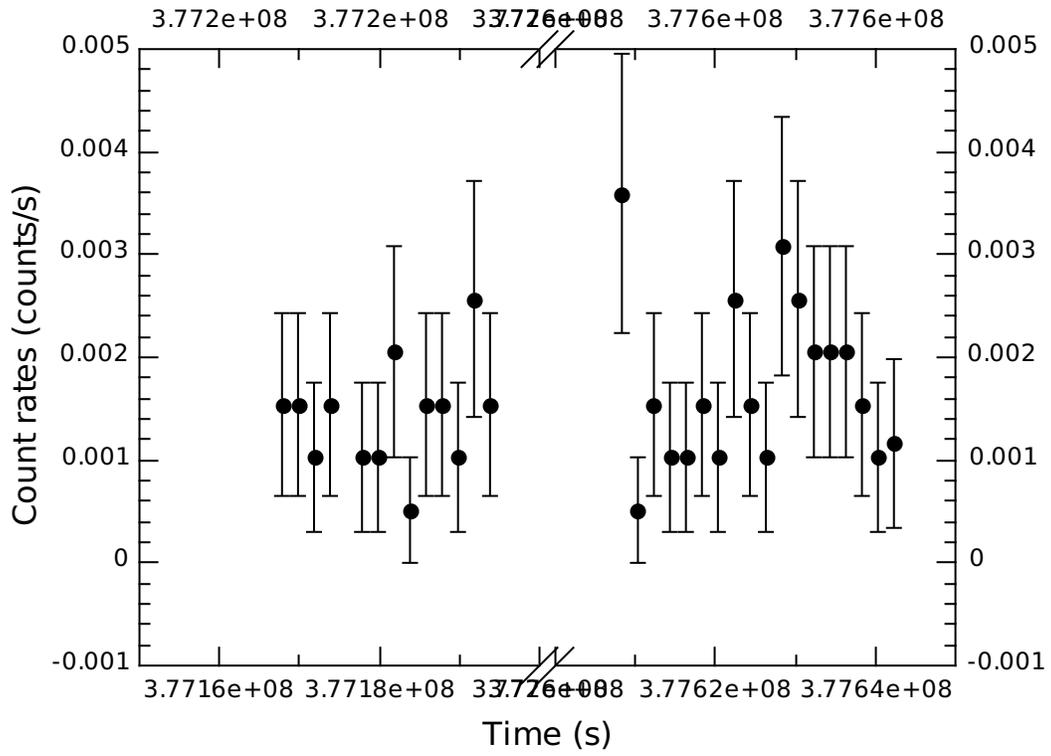}
\caption{The light curve of X1 with bin size of 2 ks obtained from the three \ch\ exposures. The time axis is broken for display purpose.
\label{f:ulx}}
\end{figure}

Ultraluminous X-ray objects (ULXs) are commonly seen in ULIRGs \citep{2003ApJ...592..782P}. They are objects that are much more luminous than a stellar mass black hole or neutron star binary accreting at Eddington limit, and are not located in the nucleus of a galaxy. As shown in Figure \ref{f:merged}, an off-nuclear point source (X1) is found about 8\arcsec\ south of the southern nucleus, which is also detected by {\tt wavdetect}. No counterpart at this position is found in optical images. We plot the light curve of X1 in Figure \ref{f:ulx} within the three \ch\ exposures, and the bin size is 2 ks. The spectra of the non-nuclear point source were extracted from a circle of 2\arcsec\ radius, and the background region was chosen to be a concentric annulus with an inner radius of 4\arcsec\ and an outer radius of 5\arcsec. This point source has 78 photon counts in the 0.3-8 keV band detected in three exposures. Thus we binned the summed spectrum into at least 1 count per bin using {\it Cash} statistics to perform the spectral fit. 

We fitted the spectrum of this point source with several models: simple power-law (PL), a thermal plasma (APEC) plus a power-law and a multi-color accretion disk (MCD) with a power-law. Table \ref{t:ulx} lists the best-fitting parameters and the observed soft and hard X-ray luminosity of each model. The best-fitting photon index is about $\Gamma=2$ in the PL and APEC+PL model, and a little bit harder ($\Gamma=1.6$) in the MCD+PL model. The temperature of the plasma in the APEC model is found to be $kT=0.3$ keV, and the temperature at inner disk radius of the MCD model is $kT_{\rm in}=0.46$ keV. All three models give the observed soft X-ray luminosity of about $2.4\times 10^{40}~{\rm ergs}~{\rm s}^{-1}$, and the hard X-ray luminosity of about $4\times 10^{40}~{\rm ergs}~{\rm s}^{-1}$, which is about one order of magnitude more luminous than a typical ULX.

\section{Discussion \label{discussion}}

The \ch\ observation allows us to disentangle the two nuclei in the merger system of IRAS 19245--7245 for the first time in X-ray imaging. Our analysis demonstrates that the southern nucleus, which hosts a Compton-thick AGN, contributes over 80\% of the X-ray emission of the entire merger system in 0.5-10 keV band, and produces $\sim 90\%$ of the hard X-ray luminosity. Extended soft X-ray emission has been observed around the southern nucleus out to a radius of $\sim$ 25 kpc, contributing a similar amount of soft X-ray photons as the southern nucleus itself.

\subsection{Emission from AGN}

{\it Soft emission:} The flat hard X-ray continuum and the intrinsic Compton-thick absorber have been well discussed by \cite{2003A&A...398..107B,2009A&A...504...53B}. One of the key goals of our \ch\ observation was to better investigate the relative contribution of SB and AGN activity. However, this discrimination becomes vague when referring to the soft emission from the southern nucleus. The soft emission is fit well with a power-law and a thermal component. The properties of the thermal component are consistent with those from the extended region and thus is likely to be thermal and starburst dominated emission rather than AGN-photoionized plasma. Even when we extracted the emission from a smaller region, this thermal component could not be ruled out. The total observed 0.5-2 keV luminosity of the southern nucleus is $2.6\times 10^{41}~{\rm ergs~s^{-1}}$, and one third comes from the thermal emission according to the spectral model. If we convert the soft X-ray luminosity to SFR assuming the relationship given by \cite{2003A&A...399...39R}, we get ${\rm SFR=57~M_{\odot}~yr^{-1}}$, which is close to the estimated SFR that is about ${\rm 80~M_{\odot}~yr^{-1}}$ for the southern nucleus \citep{2003A&A...403..119B} given the dispersion in the X-ray/SFR correlation.

{\it Fe K$\alpha$ line:} Iron emission lines are only detected in the southern nucleus. We tried to add a Gaussian component in the spectral fitting of the inner region D1 in the extended starburst region, however, it did not reduce the $\chi^2$. The low quality of the hard X-ray photons in the outer region D2 prevents us from investigating the iron line feature. Indeed, the 6-7 keV image shows that iron line photons are only from the southern nucleus. The luminosity of the iron emission line observed by \xmmn\ in March 2001 is $3.3\times 10^{41}~{\rm ergs~s^{-1}}$ (after $K-$correction), and it dropped to $2.2\times 10^{41}~{\rm ergs~s^{-1}}$ in the \suzaku\ observation in November 2005. The result from our recent \ch\ observation is, however, consistent with the \xmm\ observation. The EW of line was 1.5 keV and 0.7 keV in \xmm\ and \suzaku\ observations, respectively. 

We also investigated the possibility of a blend of two lines of the Fe K$\alpha$ emission. In the 2001 \xmm\ observation, a neutral iron line with centroid line energy at 6.4 keV and an ionized H-like iron line with centroid line energy at 6.9 keV are clearly separated at significance level $>99\%$. The intensity of the 6.4 keV line is larger than the 6.9 keV line. However, the significance of the 6.4 keV line in the 2005 \suzaku\ and 2009 \ch\ observations decreases to only $2\sigma$, and a He-like iron line (6.7 keV) shows up in the late two observations and dominates the intensity. \cite{2000ApJ...535L..79T} reported a similar iron line variability in NGC 4579 in two {\it ASCA} observations, but \cite{2002ApJ...565..108E} compared that with a \ch\ observation and argued that such variability might be due to the very large extraction region of {\it ASCA} spectra, which was contributed by the ionized Fe K$\alpha$ emission from the entire circumnuclear nebula. However, as we show in above sections that there is little extended emission in 6-7 keV band, the aperture effect is not likely the cause of the variability. 

The variability of the 6.4 keV line indicates that the cold matter where this line originates lies within 2 pc from the nucleus. If the energy shift of the ionized line from 6.9 keV to 6.7 keV is true, there might be multiple locations of the clumpy clouds with different ionization states. Since there is no flux variability of the continuum emission, we could not constrain the distance of the ionized line emitters based on the time lag between the continuum and the Fe line ionization change. If the change of ionization state (from H-like to He-like) is not true but due to the insufficient photons in the spectral fitting, there is still a $\sim 70\%$ change in the ionized Fe K$\alpha$ line intensity. The 6.7 keV line could originate either from an ionized accretion disk or from reflection/scattering of the hard nuclear X-rays by ionized clouds not farther than 2 pc from the central engine. 

\subsection{Emission from Starburst}

{\it Soft emission:} From the soft X-ray image of IRAS 19245--7245, we can see that hot gas emission extends to a radius of $\sim 20\arcsec$ ($\sim$ 25 kpc) surrounding the southern nucleus. The thermal emission requires a $kT \sim 0.8$ keV component in the best-fitting models of the inner region (D1) of the diffuse emission. The temperature is consistent with the thermal emission found in the southern nucleus. Although the spectra in the outer region (D2) have relatively poorer quality, the temperature of the hot gas in D2 from the spectral fit is consistent with that of D1 but with larger uncertainty.

Using the parameters of the X-ray spectral model, we can derive the basic physical properties of the hot gas in the regions of S, D1 and D2. The normalization for the {\tt APEC} component ($10^{-14}\int n_{e}n_{H}dV/4\pi[D_{A}(1+z)]^{2}$) implies a volume integral of density squared if we assume the electron density ($n_{e}$) approximately equals the H atom density ($n_{H}$), where $D_{\rm A}=243.3$ Mpc is the angular distance of IRAS 19254--7245 at redshift $z=0.062$. The mass ($M$), mean thermal pressure ($P$), total thermal energy ($E$) of the hot gas can be derived using the electron density and temperature from the spectral fit. The characteristic timescale could be calculated as the size divided by the sound speed in the hot gas. Then, the outflow rates of the mass ($\dot{M}$) and energy ($\dot{E}$) can be roughly estimated according to the timescale. These physical parameters are listed in Table \ref{t:physics}, where we parameterize them by the volume filling factor $f$ of the X-ray-emitting material. We see that while most of the mass and thermal energy content of the hot gas is contained in the outer regions (D1 and D2), the actual estimates of the rates at which energy and mass flow outwards are consistent as we move from the central region (S) to D1 and through D2. This is constituent with a mass- and energy-conserving outflow which originates in the starburst (inside region S).

We then use Starburst99 \citep{1999ApJS..123....3L} to estimate the relevant parameters expected for the collection of supernovae and massive stars in the starburst. We adopt a star formation rate of 80 M$_{\odot}$ yr$^{-1}$ \citep{2003A&A...403..119B} which has been constant over the past 100 Myr and has a standard Kroupa IMF. The resulting total mechanical power of the stellar wind and supernovae ejecta is $4\times 10^{43}~{\rm erg~s^{-1}}$, they carry a momentum flux of $3\times 10^{35}~{\rm dyn}$, and the mass outflow rate is $\dot{M}_{\rm SN}\sim 16~M_{\odot}~{\rm yr^{-1}}$. 

These values imply that a starburst-driven outflow can account for the properties of the extended soft X-ray emission. The observed outflow rate of (thermal) energy is only 10\% of the total energy injection rate from the starburst while the observed mass outflow rate is about three times larger than the expected starburst mass injection rate. This latter result is commonly observed in starburst-driven winds and it ascribed to mass-loading as the supernovae and massive star ejecta mix with ambient interstellar gas \citep[e.g.][]{2009ApJ...697.2030S}. The predicted momentum flux from the starburst results in a wind ram pressure of about $1.8 \times 10^{-10}$, $1.6 \times 10^{-11}$, and $4 \times 10^{-12}$ dyne cm$^{-2}$ at the outer boundaries of regions S, D1, and D2 respectively. These values are very close to the inferred pressures in the observed extended soft X-ray region.

\begin{figure}
\centering
\begin{tabular}{c}
\epsfig{file=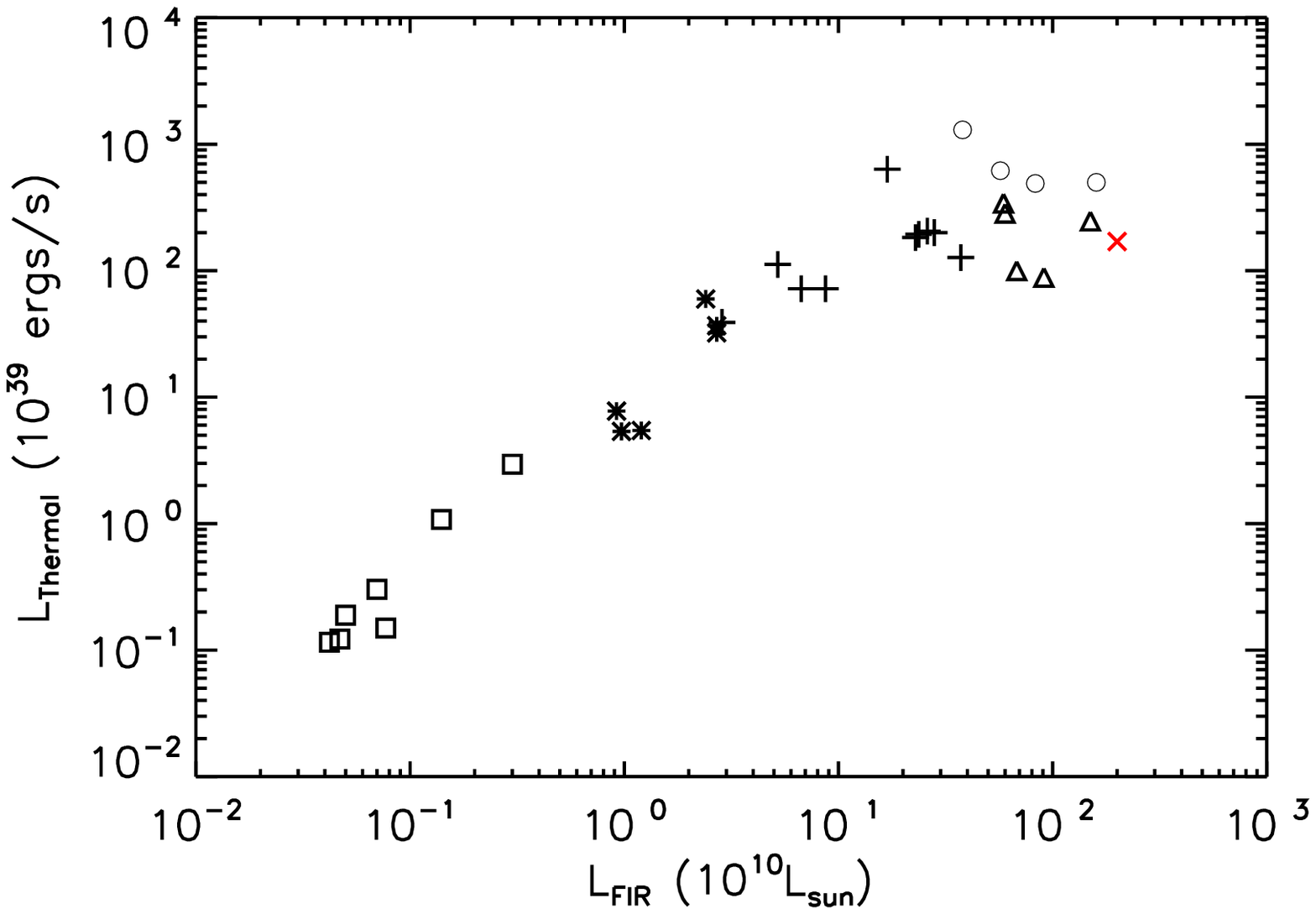,width=0.8\linewidth,clip=} \\
\epsfig{file=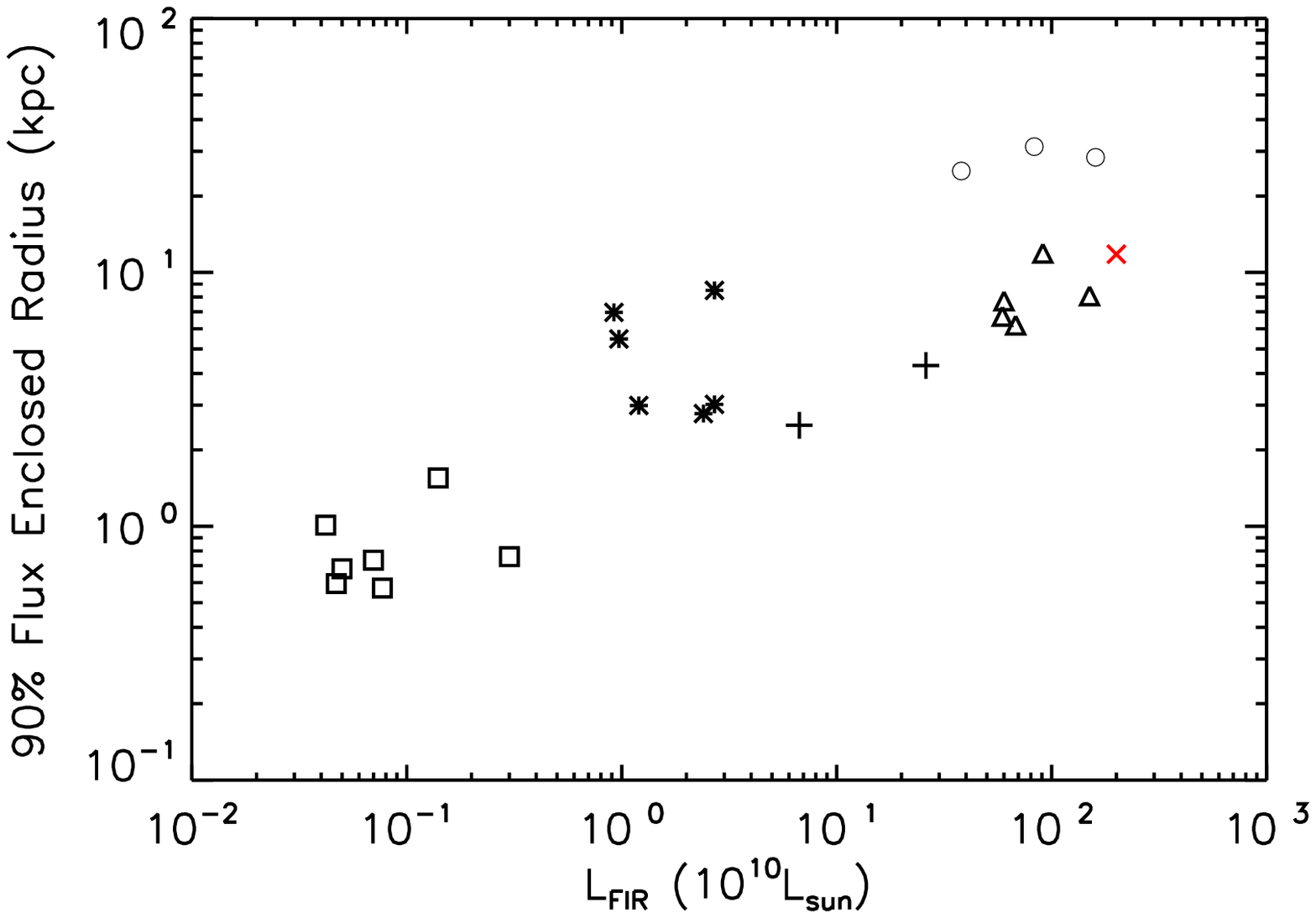,width=0.8\linewidth,clip=} \\
\end{tabular}
\caption{The total thermal X-ray luminosity ({\it upper}) and the radius enclosing 90\% of the total flux in the 0.3-2.0 keV X-ray band ({\it lower}) vs. the far-IR luminosity. IRAS 19254--7245 is marked in red cross symbol. Other sources in this plot are dwarf starbursts (square), starbursts (asterisk), Lyman break analogs (plus), starburst-dominated ULIRGs (triangle) and AGN ULIRGs (circle), which are taken from \cite{2005ApJ...628..187G,2006ApJ...648..310G,2007ApJ...668..891G,2011ApJ...731...55J}.
\label{f:fir_sx}}
\end{figure}

More empirically, we can compare the basic properties of the soft extended emission to what is seen in typical starbursts. First, we can compare the relation between the far-IR and soft X-ray luminosity of IRAS 19245--7245 with other star-forming galaxies which differ in their size and luminosity. The plot is shown in the upper panel of Figure \ref{f:fir_sx}, where the other galaxies are dwarf starbursts, normal starbursts, Lyman break analogs, starburst-dominated ULIRGs and AGN ULIRGs \citep{2005ApJ...628..187G,2006ApJ...648..310G,2007ApJ...668..891G,2011ApJ...731...55J}. The soft X-ray luminosities are derived from the thermal component in the spectral fits by excluding the contribution from AGNs and X-ray binaries. The luminosity ratio of SX/IR for ULIRGs is found to be about one dex smaller than the value for star-forming galaxies with $L_{\rm IR}<10^{11}L_{\odot}$. This deficiency in systems with high IR luminosity was characterized as $<\log L_{\rm SX}/L_{\rm IR}>=-4.55$ with $1\sigma$ scatter of 0.37 by \cite{2011MNRAS.417.2239S}. This ratio for IRAS 19245--7245 is -4.66 according to our spectral analysis, well consistent with their result for the star-forming galaxies with similarly high far-IR luminosities. We also show in the lower panel of Figure \ref{f:fir_sx} that IRAS 19245--7245 obeys the relation between the size of the soft X-ray emitting region and the far-IR luminosity defined by other star-forming galaxies. 

We conclude that the properties of the large ($\sim$ 50 kpc diameter) spatially-extended region of soft X-ray emission are entirely consistent with an origin as a starburst-powered galactic wind.

\begin{figure}
\centering
\epsfig{file=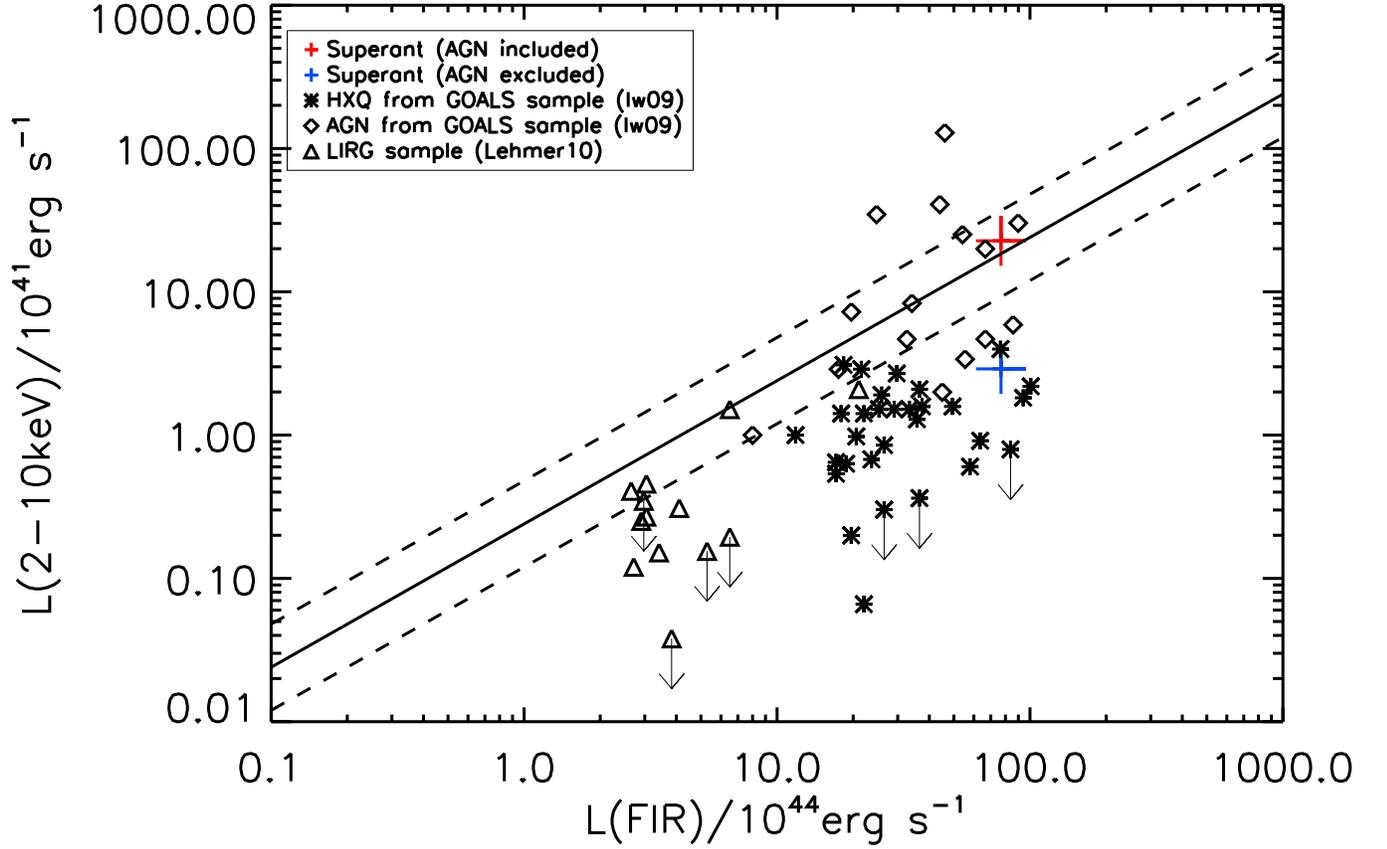}
\caption{Hard X-ray luminosity vs. far infrared luminosity. IRAS 19254--7245 is marked by plus sign in red (entire galaxies including the AGN) and in blue (without the AGN). Others are from \citet[][Iw09]{2009ApJ...695L.103I} and \citet[][Lehmer10]{2010ApJ...724..559L} for comparison, where the diamond symbols indicate the Iw09 AGN samples, and asterisks and upper limit arrows are ``Hard X-ray Quiet" galaxies (HXQs) from Iw09, and LIRG samples from Lehmer10 are marked as triangles\label{f:fir_hx}}
\end{figure}

{\it Hard emission:} The hard X-ray continuum of the diffuse emission is characterized as a power-law with photon index of $\Gamma \sim1.8$, which is much steeper than that found in southern nucleus. We compare the luminosity ratio of hard X-ray to far IR with other LIRGs and ULIRGs observed by \ch\ in Figure \ref{f:fir_hx}. The other samples are from \cite{2009ApJ...695L.103I} and \cite{2010ApJ...724..559L}, where \citeauthor{2009ApJ...695L.103I}'s sample was divided into two groups of AGNs and ``Hard X-ray Quiet" galaxies (HXQs) indicated by diamond and asterisk symbols, respectively, and \citeauthor{2010ApJ...724..559L}'s sample was indicated by triangle symbols. We show IRAS 19254--7245 as plug signs in two positions in this plot: one with AGN contribution included (in red) and the other one without AGN contribution (in blue). We see that the hard X-ray luminosity produced by the star formation process is consistent with other pure star-forming galaxies (with no AGN) having similar far-IR luminosities.

\subsection{Emission from other regions}

{\it Northern nucleus:} The northern nucleus was not resolved in the \xmmn\ observation, and it was included in the 15\arcsec\ extraction circle of spectral analysis \citep{2003A&A...398..107B}. According to our \ch\ observation, both the soft and hard X-ray luminosities are $\sim 30$ times smaller than those of the southern nucleus. We fitted the spectrum of northern nucleus with only an absorbed power-law due to the limited photons we obtained. The photon index is found to be $\Gamma=1.25$ but poorly constrained. It has been demonstrated by the optical spectra that there is no AGN emission in this nucleus. This classification is supported by its low X-ray luminosity and the non-detection of Fe K$\alpha$ line emission in our \ch\ observation. The SFR of the northern nucleus is only $\sim {\rm 0.5~M_{\odot}~yr^{-1}}$, and it appears to be in a post-starburst phase \citep{2003A&A...403..119B}, which is consistent with its weak X-ray emission.

{\it Off-nuclear point source:} One off-nuclear point source is found $\sim 8\arcsec$ south of the southern nucleus. If we assume that it is associated with the Superantennae, it would lie about 9.4 kpc from the southern nucleus in projection and have a 0.5-10 keV luminosity as high as $6\times 10^{40}~{\rm ergs~s^{-1}}$ (based on the spectral fits using different models). This is about one order of magnitude more luminous than the typical ULXs found in other galaxies, but still within the luminosity range they span. In a recent \ch\ survey of X-ray point sources in nearby galaxies, \cite{2011ApJS..192...10L} found 479 ULXs with $L_{X}(\rm 0.3-8~keV)\ge 2\times 10^{39}~{\rm erg~s^{-1}}$, and about 15\% (74 out of 479) are extreme ULXs with $L_{X}(\rm 0.3-8~keV)\ge 10^{40}~{\rm erg~s^{-1}}$.  The X-ray luminosity of this ULX in the Superantennae would imply a minimum black hole mass of about 500 M$_{\odot}$ for a source at the Eddington limit. 

We note that an association with the Supernantennae is likely. Using the 2-8 keV flux of $\sim 4\times 10^{-15}~{\rm erg~cm^{-2}~s^{-1}}$ and the number-count relations from \ch\ Deep Fields \citep{2004AJ....128.2048B,2005ARA&A..43..827B}, we calculate that the possibility of a background AGN located within a radius of 8\arcsec\ from the southern nucleus is only about $\sim 1.5\%$. 

\section{Summary}

In this work we have presented \ch\ observation of the proto-typical merger/ULIRG IRAS 19254--7245 (the Superantennae). This is the first X-ray observation that could spatially resolve the diffuse emission of this ULIRG. The main results are summarized as follows:
\begin{enumerate}
\item The southern nucleus hosts a Compton-thick AGN, which contributes $\sim 90\%$ of the hard X-ray emission of IRAS 19254--7245. The soft X-ray emission from the AGN region is contaminated by the circumnuclear starburst.

\item The diffuse emission from starburst extends out to a radius of about 25 kpc surrounding the southern nucleus, and is dominated by thermal emission with a temperature of $\sim 0.8$ keV. This emission accounts for $\sim 33\%$ of the soft emission from the whole merging system.

\item The global properties of the extended thermal emission are consistent with winds seen in many starbursts and the basic physical properties are fully consistent with an outflow driven by the supernovae and massive stars in the starburst. Feedback from the AGN is not necessary to explain what we see.

\item The prominent Fe K$\alpha$ emission line is a blend of both neutral and ionized lines, and variability is detected for the neutral one.

\item No evidence for an AGN in the northern nucleus is detected in this \ch\ observation. 

\item An off-nuclear point source is found and could be classified as a ULX if it is associated with the Superantennae.
\end{enumerate}

\acknowledgments

{}

\clearpage
\begin{table}[ht]
\setlength{\tabcolsep}{1pt}
\centering
\caption{Observation log.\label{t:summary}}
\begin{tabular}{l c c}
\hline\hline
Obs ID & Date & Exp. Time$^{\rm{a}}$ \\
& YYYY-MM-DD & (ks) \\
\hline
10538 & 2009-12-14 & 29.06\\
11797 & 2009-12-19 & 35.16\\
11798 & 2009-12-19 & 11.81\\
\hline
\tablenotetext{\rm{a}}{The exposure time is the good time interval after flare filtering.}
\end{tabular}
\end{table}

\clearpage
\begin{landscape}
\begin{deluxetable}{lccccccccccc}
\tablewidth{0pt}
\tabletypesize{\scriptsize}
\tabletypesize{\footnotesize}
\setlength{\tabcolsep}{1pt}
\tablecaption{X-ray spectral model parameters.\label{t:fit}}
\startdata
\hline\hline
Region & $N_{\rm H}$\tablenotemark{a} & $kT$\tablenotemark{b} & $\Gamma$\tablenotemark{c} & $E_{\rm Fe}$\tablenotemark{d} & $\sigma$\tablenotemark{e} & EW\tablenotemark{f} & $\chi^2/\nu$ & $L_{\rm SX,th}$\tablenotemark{g} & $L_{\rm SX,pl}$\tablenotemark{h} & $L_{\rm HX}$\tablenotemark{i} & $L_{\rm Fe}$\tablenotemark{j} \\
\hline
Global \tablenotemark{k} & ~~$1.84_{-1.43}^{+2.75}$~~ & ~~$0.72_{-0.14}^{+0.12}$~~ & ~~$1.30_{-0.16}^{+0.18}$~~ &  ~~$6.68_{-0.08}^{+0.14}$~~ & ~~$0.13_{-0.09}^{+0.17}$~~ & ~~$1.71_{-0.74}^{+0.90}$~~ & ~~146.0/154~~ & ~~$1.3_{-0.7}^{+2.1}$~~ & ~~$2.7_{-0.7}^{+0.8}$~~ & ~~$22.6_{-5.5}^{+7.0}$~~ & ~~$4.1_{-1.8}^{+2.1}$~~ \\ 
Global \tablenotemark{l} & ~~$1.29_{-0.45}^{+0.77}$~~ & ~~$0.77_{-0.07}^{+0.06}$~~ & ~~$1.26_{-0.07}^{+0.09}$~~ & $6.68_{-0.05}^{+0.05}$ & $0.17_{-0.05}^{+0.06}$ & ~~ & ~~388.9/361~~ & ~~$1.7_{-0.4}^{+0.8}$~~ & ~~$3.6_{-0.2}^{+0.4}$~~ & ~~$22.7_{-1.3}^{+2.4}$~~ \\
S (joint) & ~~$3.35_{-1.22}^{+1.57}$~~ & ~~$0.79_{-0.23}^{+0.39}$~~ & ~~$1.28_{-0.17}^{+0.25}$~~ & ~~$6.68_{-0.11}^{+0.12}$~~ & ~~$0.23_{-0.23}^{+0.50}$~~ & ~~$1.52_{-0.73}^{+1.01}$~~ & ~~81.7/103~~ & ~~$0.7_{-0.2}^{+0.4}$~~ & ~~$2.1_{-0.2}^{+0.6}$~~ & ~~$19.4_{-2.1}^{+5.3}$~~ & ~~$3.3_{-1.2}^{+1.4}$~~ \\
S (summed) & $5.26_{-2.63}^{+2.87}$ & $0.79_{-0.19}^{+0.29}$ & $1.22_{-0.21}^{+0.19}$ & $6.69_{-0.12}^{+0.22}$ & $0.24_{-0.13}^{+0.29}$ & $1.50_{-0.60}^{+0.44}$ & 99.1/97 & $0.9_{-0.5}^{+1.1}$ & $1.7_{-0.3}^{+0.3}$ & $19.8_{-3.3}^{+3.9}$ & $3.3_{-1.3}^{+1.0}$ \\
N & $2.49_{-2.49}^{+7.04}$ & & $1.25_{-1.23}^{+0.93}$ & & & & 25.5/29 & & $0.10_{-0.04}^{+0.05}$ & $0.48_{-0.19}^{+0.25}$ & \\
D1 & $0.69_{-0.69}^{+0.94}$ & $0.78_{-0.07}^{+0.13}$ & $1.73_{-0.56}^{+0.60}$ & & & & 23.1/26 & $0.58_{-0.24}^{+0.69}$ & $0.4_{-0.18}^{+0.26}$ & $0.88_{-0.41}^{+0.59}$ & \\
D2 & $N_{\rm H,G}$ & $0.90_{-0.27}^{+0.46}$ & $2.70_{-0.88}^{+0.87}$ & & & & 28.5/28 & $0.17_{-0.11}^{+0.11}$ & $0.25_{-0.17}^{+0.15}$ & $0.15_{-0.10}^{+0.09}$ & \\
D1+D2 & $1.10_{-1.02}^{+0.98}$ & $0.77_{-0.11}^{+0.08}$ & $1.79_{-0.59}^{+0.75}$ & & & & 53.8/55 & $0.86_{-0.30}^{+0.67}$ & $0.49_{-0.22}^{+0.35}$ & $1.15_{-0.52}^{+0.82}$ & \\
\enddata
\tablecomments{Uncertainties correspond to the 90\% confidence level for one interesting parameter.}
\tablenotetext{a}{Column density in units of $10^{21}$cm$^{-2}$.}
\tablenotetext{b}{Temperature of thermal plasma in keV.}
\tablenotetext{c}{Photon index of hard continuum.}
\tablenotetext{d}{Energy of line center in keV in rest-frame.}
\tablenotetext{e}{Emission line width in keV.}
\tablenotetext{f}{Equivalent width of emission line in keV.}
\tablenotetext{g}{Soft X-ray luminosity contributed by thermal plasma in units of $10^{41}$ergs s$^{-1}$.}
\tablenotetext{h}{Soft X-ray luminosity contributed by power-law continuum in units of $10^{41}$ergs s$^{-1}$.}
\tablenotetext{i}{Hard X-ray luminosity in units of $10^{41}$ergs s$^{-1}$.}
\tablenotetext{j}{Iron emission line luminosity in units of $10^{41}$ergs s$^{-1}$.}
\tablenotetext{k}{Using stacked spectral data from \ch\ observation.}
\tablenotetext{l}{Using \ch, \xmm\ and \suzaku\ data.}
\end{deluxetable}
\clearpage
\end{landscape}

\begin{deluxetable}{lccccccc}
\tablewidth{0pt}
\tabletypesize{\scriptsize}
\tabletypesize{\footnotesize}
\setlength{\tabcolsep}{1pt}
\tablecaption{Spectral fits parameters of the diffuse emission in 0.5-3 keV.\label{t:sb}}
\startdata
\hline\hline
Region & ~~~$N_{\rm H,1}$~~~ & ~~~~$kT_{1}$~~~~ & ~~~$K^{*}$~~~ & ~~~$N_{\rm H,2}$~~~ & ~~~~$kT_{2}$~~~~ & ~~~$K^{*}$~~~ & ~~~$\chi^2/\nu$~~~ \\
& $10^{21}$cm$^{-2}$ & keV & & $10^{21}$cm$^{-2}$ & keV &\\
\hline
\multicolumn{8}{c}{Single temperature} \\
\hline
D1 & $N_{\rm H,G}$ & $0.77_{-0.09}^{+0.08}$ & $4.3_{-0.8}^{+1.2}\times 10^{-6}$ & & & & 19.15/22 \\
D2 & $N_{\rm H,G}$ & $0.91_{-0.60}^{+0.48}$ & $1.2_{-1.0}^{+1.0}\times 10^{-6}$ & & & & 18.9/17 \\
\hline
\multicolumn{8}{c}{Double temperatures}\\
\hline
D1 & $21.9_{-10.2}^{+29.7}$ & $0.34_{-0.06}^{+0.23}$ & $3.5_{-3.0}^{+190}\times 10^{-4}$ & $N_{\rm H,G}$ & $0.76_{-0.09}^{+0.07}$ & $4.8_{-1.0}^{+0.8}\times 10^{-6}$ & 15.2/19 \\
D2 & $0.75^{+7.25}$ & $0.27_{-0.12}^{+0.18}$ & $2.4_{-1.8}^{+960}\times 10^{-6}$ & $3.54_{-3.54}^{+4.36}$ & $0.95_{-0.24}^{+0.45}$ & $4.0_{-3.0}^{+7.1}\times 10^{-6}$ & 18.1/13 \\
\enddata
\tablenotetext{*}{Plasma model normalization in units of $10^{-14}/\{4\pi[D_{\rm A}(1+z)]^{2}\}\int n_{e}n_{\rm H}dV$, where $D_{\rm A}$ is the angular distance, and $n_{e}$ and $n_{\rm H}$ are the electron and hydrogen number densities, respectively.}
\end{deluxetable}

\begin{deluxetable}{lcccccccccc}
\tablewidth{0pt}
\tabletypesize{\scriptsize}
\tabletypesize{\footnotesize}
\setlength{\tabcolsep}{1pt}
\tablecaption{Iron emission line features.\label{t:iron}}
\startdata
\hline\hline
& & \multicolumn{9}{c}{Spectral fit}\\
\cline{3-11}
& & \multicolumn{3}{c}{~~~~~line 1~~~~~} && & \multicolumn{3}{c}{~~~~~line 2~~~~~} &\\
\cline{3-5}\cline{8-10}
Observation~~~~~~ & & $E$ & $\sigma$ & EW && ~~~ & $E$ & $\sigma$ & EW &\\
\hline
\xmm\ & & ~~$6.40_{-0.13}^{+0.12}$~~ & ~~$0.13_{-0.09}^{+0.27}$~~ & ~~$1.16_{-0.50}^{+0.94}$~~ && & ~~$6.89_{-0.08}^{+0.07}$~~ & ~~0.01 (fix)~~ & ~~$0.44_{-0.33}^{+0.32}$~~& \\
\suzaku\ & & ~~$6.39_{-0.30}^{+0.30}$~~ & ~~0.01 (fix)~~ & ~~$0.09_{-0.06}^{+0.06}$~~ & && ~~$6.69_{-0.04}^{+0.04}$~~ & ~~$0.10_{-0.05}^{+0.06}$~~ & ~~$0.48_{-0.11}^{+0.12}$~~ &\\
\ch\ & & ~~$6.43_{-0.36}^{+0.37}$~~ & ~~0.1 (fix)~~ & ~~$0.17_{-0.17}^{+0.17}$~~ && & ~~$6.71_{-0.03}^{+0.07}$~~ & ~~$0.04_{-0.04}^{+0.09}$~~ & ~~$0.66_{-0.26}^{+0.41}$~~&\\
\hline
& & \multicolumn{9}{c}{MCMC}\\
\cline{3-11}
& & \multicolumn{4}{c}{~~~~~line 1~~~~~} & & \multicolumn{4}{c}{~~~~~line 2~~~~~}\\
\cline{3-6}\cline{8-11}
Observation~~~~~~ & & $E$ & $\sigma$ & EW & significance & ~~~ & $E$ & $\sigma$ & EW & significance \\
\hline
\xmm\ & & ~~$6.41_{-0.09}^{+0.10}$~~ &~~$0.13_{-0.08}^{+0.07}$~~& ~~$1.15_{-0.48}^{+0.54}$~~ & $>99.99\%$ & & ~~$6.89_{-0.09}^{+0.10}$~~& ~~0.01 (fix)~~ & ~~$0.44_{-0.34}^{+0.32}$~~ & $97.14\%$ \\
\suzaku\ & & ~~$6.41_{-0.13}^{+0.18}$~~& ~~0.01 (fix)~~& ~~$0.09_{-0.09}^{+0.08}$~~ & $87.79\%$ & & ~~$6.69_{-0.06}^{+0.06}$~~ &~~$0.11_{-0.03}^{+0.09}$~~ & ~~$0.47_{-0.13}^{+0.11}$~~ & $>99.99\%$ \\
\ch\ & & ~~$6.35_{-0.28}^{+0.21}$~~ & ~~0.01 (fix)~~ & ~~$0.14_{-0.14}^{+0.17}$~~ & $81.16\%$ & & ~~$6.71_{-0.05}^{+0.06}$~~ &~~$0.08_{-0.03}^{+0.02}$~~ & ~~$0.89_{-0.31}^{+0.37}$~~ & $>99.99\%$ \\
\enddata
\tablecomments{$E$, $\sigma$ and EW are in units of keV.}
\end{deluxetable}

\begin{deluxetable}{lcccccc}
\tablewidth{0pt}
\tabletypesize{\scriptsize}
\tabletypesize{\footnotesize}
\setlength{\tabcolsep}{1pt}
\tablecaption{Spectral fits parameters of the off-nuclear point source (X1).\label{t:ulx}}
\startdata
\hline\hline
model~~~~~~~ & ~~~~~~$N_{\rm H}$~~~~~~ & $kT$ or $kT_{\rm in}$ & ~~~~~~~~$\Gamma$~~~~~~~~ & $C$-statistic & $L_{\rm 0.5-2keV}$ & $L_{\rm 2-10keV}$ \\
& $\times 10^{21}~{\rm cm}^{-2}$ & keV & & /dof & $\times 10^{40}~{\rm ergs~s}^{-1}$ & $\times 10^{40}~{\rm ergs~s}^{-1}$\\
\hline
PL & $1.37_{-0.67}^{+0.81}$ && $2.03_{-0.65}^{+0.82}$ & 50.0/55 & $2.4_{-0.4}^{+0.5}$ & $4.1_{-0.8}^{+0.9}$\\
APEC+PL & $2.79_{-0.76}^{+0.79}$ & $0.29_{-0.11}^{+0.18}$ & $2.07_{-0.71}^{+0.77}$ & 49.0/53 & $2.3_{-0.5}^{+0.6}$ & $4.0_{-0.9}^{+1.1}$\\
MCD+PL & $N_{\rm H,G}$ & $0.46_{-0.14}^{+0.07}$ & $1.58_{-0.30}^{+0.38}$ & 49.7/54 & $2.4_{-1.8}^{+2.1}$ & $4.5_{-1.2}^{+1.4}$ \\
\enddata
\end{deluxetable}

\clearpage
\begin{table}[ht]
\setlength{\tabcolsep}{1pt}
\centering
\caption{Physical properties of the diffuse hot gas in regions S, D1 and D2.\label{t:physics}}
\begin{tabular}{l c c c}
\hline\hline
Parameters & S & D1 & D2 \\
\hline
$n_{e}~({\rm 10^{-2}~cm^{-3}})$ & ~~~$5.5f^{-1/2}_{0}$~~~ & ~~~$0.44f^{-1/2}_{1}$~~~ & ~~~$0.087f^{-1/2}_{2}$~~~\\
$M~(\rm 10^{8}~M_{\odot})$ & $2.5f^{1/2}_{0}$ & $7.2f^{1/2}_{1}$ & $10.3f^{1/2}_{2}$\\
$\dot{M}~(M_{\odot}~{yr}^{-1})$ & $39f^{1/2}_{0}$ & $47f^{1/2}_{1}$ & $48f^{1/2}_{2}$\\
$E~({\rm 10^{56}~erg})$ & $8.7f^{1/2}_{0}$ & $25f^{1/2}_{1}$ & $36f^{1/2}_{2}$ \\
$\dot{E}~({\rm 10^{42}~erg s^{-1}})$ & $4.3f^{1/2}_{0}$ & $5.2f^{1/2}_{1}$ & $5.3f^{1/2}_{2}$\\
$P~({\rm 10^{-10}~dyn~cm^{-2}})$ & $1.6f^{-1/2}_{0}$ & $0.13f^{-1/2}_{1}$ & $0.025f^{-1/2}_{2}$\\
\hline
\end{tabular}
\tablecomments{$f_{0}$, $f_{1}$ and $f_{2}$ are the volume filling factors of the X-ray emitting material in regions of S, D1 and D2, respectively. See text for details.}
\end{table}

\end{document}